\documentclass[prb,aps,amsmath,amssymb,floatfix,reprint]{revtex4-1}
\usepackage{graphicx}
\usepackage{bm}
\usepackage{amsfonts,mathrsfs}
\usepackage{hyperref}
\hypersetup{colorlinks,citecolor=blue,linkcolor=blue,urlcolor=blue}

\begin{document}
\title{Local Structure Analysis in \textit{Ab Initio} Liquid Water}
\author{Biswajit Santra$^{1}$}
\author{Robert A. DiStasio, Jr.$^{1}$}
\author{Fausto Martelli$^{1}$}
\author{Roberto Car$^{1,2,3,4}$}
\affiliation{
$^1$Department of Chemistry, Princeton University, Princeton, NJ 08544, USA \\
$^2$Department of Physics, Princeton University, Princeton, NJ 08544, USA \\
$^3$Princeton Institute for the Science and Technology of Materials, Princeton University, Princeton, NJ 08544, USA \\
$^4$Program in Applied and Computational Mathematics, Princeton University, Princeton, NJ 08544, USA
}

\begin{abstract}
Within the framework of density functional theory, the inclusion of exact exchange and non-local van der Waals/dispersion (vdW) interactions is crucial for predicting a microscopic structure of ambient liquid water that quantitatively agrees with experiment.
In this work, we have used the local structure index (LSI) order parameter to analyze the local structure in such highly accurate \textit{ab initio} liquid water.
At ambient conditions, the LSI probability distribution, P($I$), was unimodal with most water molecules characterized by more disordered high-density-like local environments.
With thermal excitations removed, the resultant bimodal P($I$) in the inherent potential energy surface (IPES) exhibited a 3:1 ratio between high- and low-density-like molecules, with the latter forming small connected clusters amid the predominant population.
By considering the spatial correlations and hydrogen bond network topologies \textit{among} water molecules with the same LSI identities, we demonstrate that the signatures of the experimentally observed low- (LDA) and high-density (HDA) amorphous phases of ice are present in the IPES of ambient liquid water.
Analysis of the LSI autocorrelation function uncovered a persistence time of $\sim$ 4 ps---a finding consistent with the fact that natural thermal fluctuations are responsible for transitions between these distinct yet transient local aqueous environments in ambient liquid water.
\end{abstract}

\maketitle

\section{Introduction}

Water is one of the most abundant molecules on Earth, yet the molecular arrangements in the solid, liquid, or even gas phases of water are extraordinarily complex.
In the microscopic structure of water, this complexity arises due to a subtle balance between directional hydrogen-bonded interactions and weaker van der Waals (vdW) interactions which are non-directional~\cite{Distasio2014,Stillinger1980,franks_book_2000,Santra2011,Santra2013,Santra2009,Santra2008}.
In fact, this set of intricate intermolecular interactions gives rise to many of the non-trivial thermodynamic and kinetic behaviors observed in water, particularly in the liquid phase, which significantly differs from many other simple liquids in properties such as the isothermal compressibility, heat capacity, and thermal expansivity, which all seem to diverge at low temperature~\cite{Angell1982,Mishima1998,Debenedetti2003}.
It has been argued that these anomalous properties are related to the observation of two metastable glassy ice forms, the low-density (LDA) and high-density amorphous (HDA) ices, which can be prepared experimentally and are separated by a first-order-like phase boundary~\cite{Mishima1985,Mishima1991,Mishima2002,Mishima1994,Winkel2008,Winkel2011}.
LDA and HDA recovered at low temperature ($\approx$ 80 K) and ambient pressure differ by their densities, \textit{i.e.}, $\rho_{\rm LDA}$=0.93 g/cm$^3$ and $\rho_{\rm HDA}$=1.17 g/cm$^3$~\cite{Finney2002}.
The molecular environment in LDA is more similar to that of crystalline ice (cubic or hexagonal) while that of HDA is more disordered and similar to liquid water at standard temperature and pressure (STP).
Furthermore, there is some experimental evidence for the existence of two metastable liquid states above the corresponding two glass transition temperatures~\cite{Amann-Winkel2013,Andersson2006,Andersson2011}, and these two liquid extensions of the amorphous ices are respectively called the low-density (LDL) and high-density liquids (HDL).
In this regard, it has been suggested that the coexistence line separating LDA and HDA would extend into the liquid regime terminating at a critical point called the liquid-liquid (LL) critical point (LLCP)~\cite{Poole1992}, and this scenario would provide a possible explanation for the anomalous response properties of water.
However, it has not been possible to experimentally prove two-liquid coexistence to date, since crystallization is too fast in macroscopic samples in the deeply supercooled regime.
Very recently, however, water droplets with diameters of $\approx$ 10 $\mu$m were found to remain in the liquid phase for approximately 1 ms at 1 bar and 227$^{+2}_{-1}$ K, ~\textit{i.e.}, at a temperature near the hypothesized LLCP~\cite{Sellberg2014}.
The structure factors of these liquid droplets were measured with ultra-fast x-ray techniques and found to be very close to that of LDA, as one would expect for LDL~\cite{Sellberg2014}.
While direct experimental evidence of a liquid-liquid coexistence in deeply supercooled water has been so far elusive, computer simulations have shown coexistence with certain classical model potentials (force fields)~\cite{Palmer2014,Tanaka2002} such as ST2~\cite{ST2model}, although other force fields, such as the coarse-grained mW model, do not show such behavior~\cite{Moore2010}.
At temperatures above the hypothesized LLCP, undercooled water transforms continuously from LDL to HDL with increasing pressures as shown by neutron scattering data~\cite{Soper2000}.
Short-lived local fluctuations associated to LDL or HDL environments have long been suggested to show up in experimental observations of the liquid at STP~\cite{Huang2009,Nilsson2011,Nilsson2012}.

To date, classification of the structures characterizing the LDL and HDL molecular environments has been accomplished using classical force fields in conjunction with several order parameters to differentiate the local structural environments in these two forms of liquid water~\cite{Poole1992,Mishima1998,Debenedetti2003,Palmer2014}.
In this regard, recent studies have shown that the local structure index (LSI)~\cite{Shiratani1996,Shiratani1998} is a very insightful order parameter that can be employed to characterize the LDL and HDL molecular environments~\cite{Palmer2014}.
For instance, the distribution of the LSI order parameter was found to be bimodal in the corresponding inherent potential energy surface (IPES) of both~\textit{supercooled} and~\textit{ambient} liquid water~\cite{Appignanesi2009,Accordino2011,Wikfeldt2011}.
However, no attempt has been made yet to analyze the local environment in liquid water based on structures (configurations) generated using the more sophisticated \textit{ab initio} based potentials available within the framework of density functional theory (DFT).
While the typical structural relaxation times in the supercooled regime are too long to be studied at present using DFT-based \textit{ab initio} molecular dynamics (AIMD) simulations, these techniques can be utilized to accurately analyze the diverse local structural environments in ambient liquid water, which has also been the subject of considerable debate in the recent literature~\cite{Huang2009,Clark2010,Sedlmeier2011,English2011,Nilsson2012}.

The predictive power of DFT-based AIMD simulations depends crucially upon the accuracy of the underlying exchange-correlation (XC) functional utilized in the quantum mechanical treatment of the electronic degrees of freedom.
In this regard, a number of studies have now demonstrated that the most commonly utilized XC functionals based on the generalized gradient approximation (GGA) provide an inadequate description of liquid water at ambient conditions---a description in which the predicted liquid is significantly overstructured and exhibits excessively sluggish dynamics~\cite{Distasio2014,asthagiri_pre_2003,grossman_jcp_2004,schwegler_jcp_2004,fernandez_jcp_2004,kuo_jpcb_2004,mcgrath_cpc_2005,vandevondele_jcp_2005,sit_jcp_2005,fernandez_ms_2005,mcgrath_mp_2006,lee_jcp_2006,todorova_jpcb_2006,lee_jcp_2007,guidon_jcp_2008,kuhne_jctc_2009,mattson_jctc_2009,yoo_jcp_2009,zhang_jctc_2011-2,Bartok2013,Alfe2013,lin_jpcb_2009,jonchiere_jcp_2011,wang_jcp_2011,zhang_jctc_2011,mogelhoj_jpcb_2011,lin_jctc_2012,yoo_jcp_2011,schmidt_jpcb_2009,ma_jcp_2012,ben_jpcl_2013,Corsetti2013,Bankura2014,Forster-Tonigold2014,Miceli2015}.
These limitations of GGA-XC functionals stem from the well-known and deleterious effects of self-interaction error~\cite{perdew_prb_1981} as well as the neglect of the non-local electron correlation effects responsible for van der Waals (vdW) or dispersion interactions.
As such, a reduction in the self-interaction error~\textit{via} the incorporation of a fraction of exact (Hartree-Fock) exchange and an explicit inclusion of vdW/dispersion interactions have individually and collectively yielded a more accurate description of the structural and dynamical properties of liquid water~\cite{Distasio2014,zhang_jctc_2011,zhang_jctc_2011-2,wang_jcp_2011}.

In this regard, we have recently shown that utilization of the PBE0 functional~\cite{perdew_jcp_1996,adamo_jcp_1999}, which includes 25\% exact exchange, in conjunction with a fully self-consistent (SC) implementation of the density-dependent vdW/dispersion correction of Tkatchenko and Scheffler~\cite{tkatchenko_prl_2009} (TS-vdW),~\textit{i.e.}, the PBE0+TS-vdW(SC) XC functional, yields an oxygen-oxygen structure factor, $S_{\rm OO}(Q)$, and corresponding radial distribution function, $g_{\rm OO}(r)$, that are in quantitative agreement with the best available experimental data~\cite{Distasio2014,Skinner2013}.
This level of agreement between~\textit{ab initio} simulations and experiment primarily originates from an increase in the relative population of water molecules in the interstitial region between the first and second coordination shells, a collective reorganization in the liquid phase which is facilitated by a weakening of the hydrogen bond strength by the use of the PBE0 hybrid XC functional, coupled with a relative stabilization of the resultant disordered liquid water configurations by the inclusion of non-local vdW/dispersion interactions.
This increasingly more accurate description of the underlying hydrogen bond network in liquid water also yielded other correlation functions, such as the oxygen-hydrogen radial distribution function, $g_{\rm OH}(r)$, and the higher-order oxygen-oxygen-oxygen triplet angular distribution, $P_{\rm OOO}(\theta)$, which encodes the degree of local tetrahedrality, as well as electrostatic properties, such as the effective molecular dipole moment, that are in better agreement with experiment~\cite{Distasio2014}.
Therefore, the overall agreement between experiment and the PBE0+TS-vdW(SC) description of the microscopic structure of ambient liquid water is indeed a very promising starting point for accurately measuring and further exploring the fluctuations in the local structural environments of liquid water by means of the LSI order parameter.

In this work, we have extensively analyzed this continuous distribution of local molecular environments in liquid water at ambient conditions and at the corresponding IPES using a systematic hierarchy of DFT XC functionals.
For all of the XC functionals considered herein, the probability distributions of the LSI order parameter, P($I$), were found to be unimodal in shape and rapidly decaying from maxima located in the low-LSI range, indicating that a majority of water molecules at ambient conditions are situated in more disordered high-density-like local environments.
In this regard, we found that the proportion of molecules having low- and high-density-like environments varies significantly with the choice of the underlying XC functional, a trend which essentially manifests as a temperature effect.
By removing thermal excitations and obtaining the IPES corresponding to ambient liquid water generated at the PBE0+TS-vdW(SC) level of theory, we found that the resultant P($I$) was bimodal in shape and exhibited a 3:1 ratio between high- and low-density-like molecules, with the latter forming small connected clusters amid the predominant population of high-density-like molecules.
By considering the spatial correlations and the underlying hydrogen bond network topologies \textit{among} IPES water molecules characterized by the same LSI identity, we demonstrate that distinctive signatures of the experimentally observed polymorphism in the amorphous phases of ice are also present in the IPES of ambient liquid water.
Further analysis of the temporal correlations \textit{via} the LSI autocorrelation function has also uncovered a persistence time of $\sim$ 4 ps---a finding which is consistent with the fact that LSI fluctuations in ambient liquid water occur due to natural thermal fluctuations between these distinct yet transient local aqueous environments.

The remainder of the paper is organized as follows.
In Section~\ref{sec:ComputationalMethods}, we describe the computational details of the AIMD simulations performed herein and the definition of the LSI order parameter.
Section~\ref{sec:ResultsandDiscussion} contains an in-depth discussion of the results of these simulations and a comparative analysis with the currently available theoretical and experimental literature.
Finally, brief conclusions are provided in Section~\ref{sec:Conclusions}.


\section{Computational Methods \label{sec:ComputationalMethods}}

\subsection{Simulation Details \label{subsec:SimulationDetails}}

In this work, we have systematically performed a series of Car-Parrinello AIMD simulations~\cite{car_prl_1985} of ambient liquid water using a hierarchy of different XC functionals.
The sequence of XC functionals employed herein includes the standard semi-local GGA of Perdew, Burke, and Ernzerhof (PBE)~\cite{perdew_prl_1996}, the corresponding hybrid PBE0~\cite{perdew_jcp_1996,adamo_jcp_1999} which includes 25\% exact exchange, and the self-consistent (SC) dispersion-corrected analogs~\cite{distasio_unpublished} thereof, \textit{i.e.}, PBE+TS-vdW(SC) and PBE0+TS-vdW(SC), based on the Tkatchenko-Scheffler~\cite{tkatchenko_prl_2009} density-dependent vdW/dispersion functional.

All of these AIMD simulations were performed in the canonical ($NVT$) ensemble using periodic simple cubic simulation cells with lattice parameter set to reproduce the experimental density of liquid water at ambient conditions.
All of the AIMD simulations were initially equilibrated for approximately 2 ps and then continued for at least an additional 20 ps for data collection.
Four AIMD simulations were performed on (H$_2$O)$_{64}$ at 300 K using the PBE, PBE0, PBE+TS-vdW(SC), and PBE0+TS-vdW(SC) XC functionals.
Since a classical treatment of the nuclear degrees of freedom is insufficient for a quantitatively accurate description of the microscopic structure of ambient liquid water, we have also performed an additional AIMD simulation at 330 K on (H$_2$O)$_{128}$ at the PBE0+TS-vdW(SC) level, a technique suggested by the lowest-order perturbative expansion in $\hbar$ of the free energy to account for the quantum mechanical nature of the nuclear degrees of freedom~\cite{landau_sm_book_1969}.
In practice, this increase of approximately 30 K in the simulation temperature has been found to mimic the nuclear quantum effects (NQE) in structural quantities such as the oxygen-oxygen radial distribution function ($g_{\rm OO}(r)$) in both DFT~\cite{morrone_prl_2008} and force field~\cite{paesani_jcp_2007,fanourgakis_jcp_2006} based MD simulations.

All calculations reported herein were performed within the plane-wave and pseudopotential framework and utilized a modified development version of the Quantum ESPRESSO (QE) software package~\cite{QE-2009}.
To meet the additional computational demands associated with large-scale AIMD simulations based on hybrid XC functionals, we have employed a linear scaling O($N$) exact exchange algorithm that exploits the natural sparsity associated with the real-space maximally localized Wannier function (MLWF)~\cite{marzari_prb_1997} representation of the occupied Kohn-Sham electronic states, which has been developed~\cite{wu_prb_2009} and extensively optimized~\cite{ko_unpublished} in our research group.
In addition, we have also developed and utilized a linear scaling O($N$) self-consistent implementation of the TS-vdW dispersion correction~\cite{distasio_unpublished}, which provides a framework for computing atomic $C_6$ dispersion coefficients as explicit functionals of the charge density, \textit{i.e.}, $C_{6,AB}=C_{6,AB}[\rho(\mathbf{r})]$, thereby accounting for the local chemical environment of each atom~\cite{tkatchenko_prl_2009}.
More explicit descriptions of the simulation details and theoretical methods employed herein are provided in Ref.~\cite{Distasio2014}.

For comparative purposes, we have also performed several simulations of liquid water using the TIP4P/Ice~\cite{Abascal2005} rigid force field, a classical water potential chosen because of its ability to reproduce the experimental freezing temperature of liquid water to within 1 K.
Using the DL-POLY code~\cite{Todorov2006}, all classical simulations were performed for 2--10 ns (after equilibration for 50--100 ps) on (H$_2$O)$_{256}$ in the $NVT$ ensemble with periodic simple cubic simulation cells set to reproduce the experimental density of ambient liquid water.
To control the ionic temperature, Nos\'e-Hoover chain thermostats~\cite{martyna_jcp_1992} were employed in conjunction with an integration time step of 1 fs.
The Lennard-Jones (LJ) potential was truncated at 9.5~\AA\ and all electrostatic contributions were computed using the standard Ewald summation technique~\cite{Essmann1995}.

\subsection{Local Structure Index (LSI) \label{subsec:LSI}}

To quantify the degree of inhomogeneity in the local molecular environments of liquid water, we have utilized an order parameter introduced by Shiratani and Sasai~\cite{Shiratani1996,Shiratani1998}, which associates a local structure index (LSI) to the individual water molecules comprising the liquid.
In essence, the LSI order parameter is the mean-squared-deviation among the radial distances corresponding to the set of molecules that surround a given water molecule.
To compute the LSI value for a given water molecule, the set of radial oxygen-oxygen distances $\{r_j\}$ corresponding to the $N$ neighboring molecules that are within a cutoff distance of 3.7~\AA\ from the reference molecule are first ordered as follows: $r_1 < r_2 < \cdots < r_j < r_{j+1} < \cdots < r_{N} < 3.7$~\AA\ $< r_{N+1}$.
The LSI value ($I$) is then defined as the inhomogeneity in this distribution of radial distances, \textit{i.e.},
\begin{equation}
\label{eq:lsi}
I = \frac{1}{N} \sum_{j=1}^{N} \left[ \Delta_{j+1,j}-\langle\Delta\rangle \right] ^2 ,
\end{equation}
in which $\langle\Delta\rangle$ is the mean over all $\Delta_{j+1,j} \equiv r_{j+1}-r_{j}$.
Hence, $I$ provides a convenient quantitative measure of the fluctuations in the distance distribution surrounding a given water molecule within a sphere defined by a radius of $\approx$ 3.7~\AA.
In doing so, $I$ measures the extent to which a given water molecule is surrounded by well-defined first and second coordination shells.

\begin{figure}[t!]
\begin{center}
\fbox{\includegraphics[width=8.50cm]{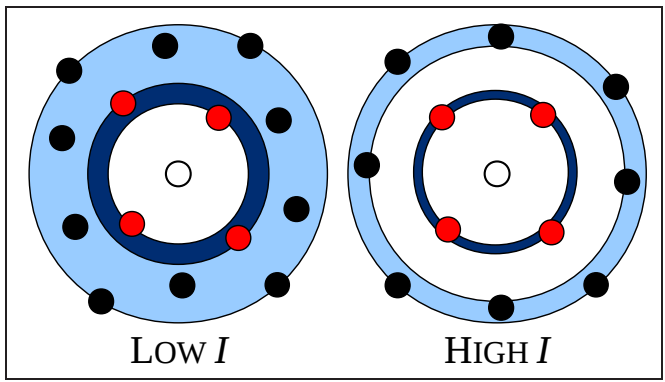}}
\caption{\label{fig:cartoon} Schematic description of high-density-like and locally disordered (\textit{left}) \textit{vs.} low-density-like and locally ordered (\textit{right}) environments, which correspond to \textit{low} and \textit{high} values of the LSI order parameter, $I$, respectively. The dark and light blue areas represent the first and second coordination shells around the central water molecule, respectively. The separation between the water molecules in the first and second coordination shells (depicted by red and black circles, respectively) are less prominent in aqueous environments characterized by a low LSI value.}
\end{center}
\end{figure}

The schematic diagram provided in Figure~\ref{fig:cartoon} illustrates the different local molecular environments that can be distinguished and quantified by the LSI order parameter.
For instance, a molecule characterized by a high $I$ value will be found in a more ordered local environment, in which the neighboring water molecules are concentrated in the region between 2.7--3.2~\AA\ and sparse in the region between 3.2--3.8~\AA.
This leads to a more prominent separation between the first and second coordination shells and a relatively low local atomic number density (as depicted on the~\textit{right} of Figure~\ref{fig:cartoon}).
On the other hand, a molecule characterized by a low $I$ value implies that the molecule is situated in a locally disordered environment, a consequence of which is a relatively high packing of neighboring molecules in the interstitial region and hence an increase in the local atomic number density (as depicted on the~\textit{left} of Figure~\ref{fig:cartoon}).

\section{Results and Discussion \label{sec:ResultsandDiscussion}}

\begin{figure*}[t!]
\begin{center}
\includegraphics[width=16cm]{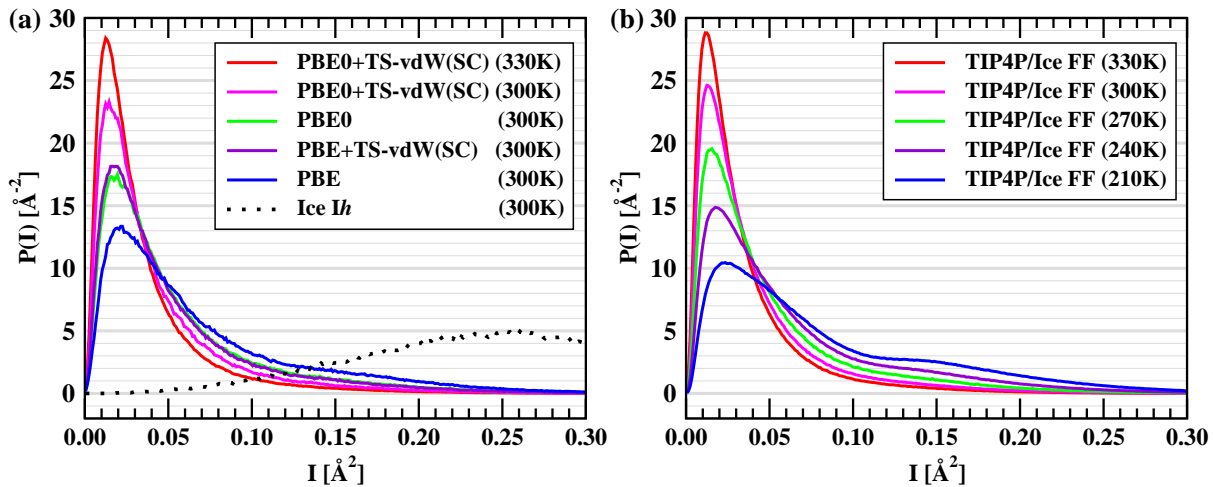}
\caption{\label{fig:plsi} Probability density distributions of the local structure index (P($I$)) in liquid water obtained from (a) DFT-based AIMD simulations and (b) classical force field (TIP4P/Ice) based simulations.}
\end{center}
\end{figure*}

\subsection{LSI Distribution in Ambient Liquid Water \label{subsec:AmbientLSI}}

The diversity of local microscopic environments found in ambient liquid water can be characterized by the probability density distribution of the LSI order parameter (P($I$)) and we begin our discussion by analyzing how P($I$) depends on the underlying DFT XC potential and the simulation temperature.
As shown in Figure~\ref{fig:plsi}(a), the P($I$) in ambient liquid water obtained from various DFT-based AIMD simulations are all unimodal in shape and rapidly decaying from maxima located between 0.013~\AA$^2$ (PBE0+TS-vdW(SC) at 330 K) and 0.023~\AA$^2$ (PBE at 300 K).
This finding is indicative of the fact that a majority of the molecules are situated in high-density-like locally disordered environments in ambient liquid water.
To illustrate that the smaller LSI values indeed correspond to more disordered local environments, we have computed P($I$) for the tetrahedrally-ordered ice I\textit{h} phase at 300 K, which is quite broadly distributed around the relatively larger LSI value of 0.25~\AA$^2$ and vanishes at $I=$ 0.05~\AA$^2$ (Figure~\ref{fig:plsi}(a)).
Considering the most accurate XC functional employed herein, PBE0+TS-vdW(SC), approximately 80\% of water molecules were found with $I<$ 0.05~\AA$^2$ at 330 K---a simulation temperature that has been elevated by 30 K to approximately account for NQE in ambient liquid water.
In this regard, one can immediately notice that a 30 K decrease in the simulation temperature (while keeping the underlying XC functional fixed) leads to a decrease of approximately 8\% in the number of water molecules that are situated in disordered high-density-like local environments (\textit{i.e.}, as characterized by $I<$ 0.05~\AA$^2$).
This is accompanied by an increase in the relative population of water molecules with more ordered low-density-like local environments,~\textit{i.e.}, water molecules characterized by larger LSI values.

To further explore the temperature dependence of P($I$), we have employed the classical TIP4P/Ice potential, a rigid water force field which is able to reproduce the experimental freezing point of liquid water to within 1 K~\cite{Abascal2005}.
As shown in Figure~\ref{fig:plsi}(b), by systematically lowering the simulation temperature from 330 K to 210 K in increments of 30 K (fixing the simulation cell corresponding to the experimental density of ambient liquid water), we observed a systematic decrease in the relative population of water molecules with lower LSI values (\textit{i.e.}, molecules in locally disordered environments) coupled with an increase in the relative population of water molecules with higher LSI values (\textit{i.e.}, molecules in locally ordered environments).
Similar to the~\textit{ab initio} PBE0+TS-vdW(SC) simulations discussed above, the relative population of water molecules with $I<$ 0.05~\AA$^2$ decreases by approximately 5\% when the simulation temperature of the liquid is reduced from 330 K to 300 K using the TIP4P/Ice force field.
Here, we note in passing that the dependence of P($I$) on temperature has also been demonstrated with other classical water force fields~\cite{Appignanesi2009,Accordino2011,Wikfeldt2011}, yielding similar qualitative results to those observed in this work.

Quite remarkably, the dependence of P($I$) on the underlying DFT XC potential essentially manifests as a temperature effect.
As shown in Figure~\ref{fig:plsi}(a), the LSI distributions obtained with the PBE and PBE+TS-vdW(SC) XC functionals are markedly different from that obtained with PBE0+TS-vdW(SC), despite the fact that the simulation temperature was set to 300 K in all of these cases.
For example, the peak height in the P($I$) computed with PBE is $\approx$ 43\% smaller than that obtained at the PBE0+TS-vdW(SC) level of theory at 300 K.
Consequently, only 50\% of the PBE water molecules were characterized by $I<$ 0.05~\AA$^2$ as opposed to 72\% in liquid water generated with PBE0+TS-vdW(SC) at 300 K.
These differences clearly indicate that the molecules in PBE-generated liquid water sample high-density-like disordered local environments to a much lesser extent than in liquid water obtained using more accurate vdW-inclusive hybrid XC functionals.
As such, this enhanced population of molecules having more ordered low-density-like local structures in ambient PBE liquid water can readily explain the overstructured pair correlation functions and exceedingly sluggish dynamics routinely observed at the PBE level of theory~\cite{grossman_jcp_2004,schwegler_jcp_2004,fernandez_jcp_2004,sit_jcp_2005,vandevondele_jcp_2005,yoo_jcp_2009}.

To follow up on this point, it is noteworthy to reiterate the fact that the variations in P($I$) with the choice of the underlying XC potential show a significant resemblance to the systematic reductions observed in the peak heights of P($I$) while supercooling liquid water using the classical TIP4P/Ice potential.
In this regard,~\textit{ab initio} simulations employing the more accurate PBE0+TS-vdW(SC) XC functional at 330 K show a remarkable similarity with the TIP4P/Ice simulation at $\approx$ 330 K~\cite{rdf_footnote}, whereas the P($I$) obtained at the PBE level of theory more closely resembles the P($I$) distributions generated by the classical TIP4P/Ice simulations performed between 210--240 K.
From this observation, one can infer that PBE liquid water simulated at ambient temperature ($\approx$ 300 K) behaves more like a deeply supercooled liquid, which is~\textit{effectively} 60--90 K lower than the actual simulation temperature, a finding which is in line with the previous independent estimate of the melting temperature of ice with the PBE XC functional, found to be approximately 100 K higher than the experimental value~\cite{yoo_jcp_2009}.

Over the past decade, liquid water generated by the PBE XC functional at ambient conditions has been referred to as ``supercooled'' due to the following characteristic features, which include overstructuring~\cite{fernandez_jcp_2004}, underestimated diffusion coefficient~\cite{sit_jcp_2005}, and overestimated melting temperature~\cite{yoo_jcp_2009}.
From the analysis of P($I$) above, the sluggishness associated with PBE liquid water can physically be explained by this relatively increased population of low-density-like molecules---a population which is energetically stabilized by the favorable hydrogen bonding motifs that are facilitated by the presence of a locally ordered environment.
Since the fluidity in liquid water arises from the fact that the hydrogen bonds comprising the underlying tetrahedral network are continuously breaking and forming, the relatively large number of intact hydrogen bonds in this population of low-density-like water molecules leads to an exceedingly sluggish liquid characterized by a significantly underestimated diffusion coefficient.
On the other hand, the formation and stabilization of an increased population of high-density-like water molecules---as found in the AIMD simulations that employ vdW-inclusive hybrid XC functionals---are primarily driven by entropic effects~\cite{Holten2012}, which lead to a decrease in the average number of intact hydrogen bonds per water molecule and therefore a significantly more diffusive liquid.
This increasing disparity between liquid water generated at the PBE and PBE0+TS-vdW(SC) levels of theory will become even more apparent when the LSI distribution is computed from the inherent potential energy surface (IPES)~\cite{StillingerJCP1984,Stillinger1984,Stillinger1988} of ambient liquid water,~\textit{i.e.}, when the thermal fluctuations present in the liquid are removed, which brings us to the subject of the next section.

\subsection{LSI Distribution in the Inherent Potential Energy Surface (IPES) of Ambient Liquid Water \label{subsec:InherentLSI}}

\begin{figure}[t!]
\begin{center}
\includegraphics[width=8.75cm]{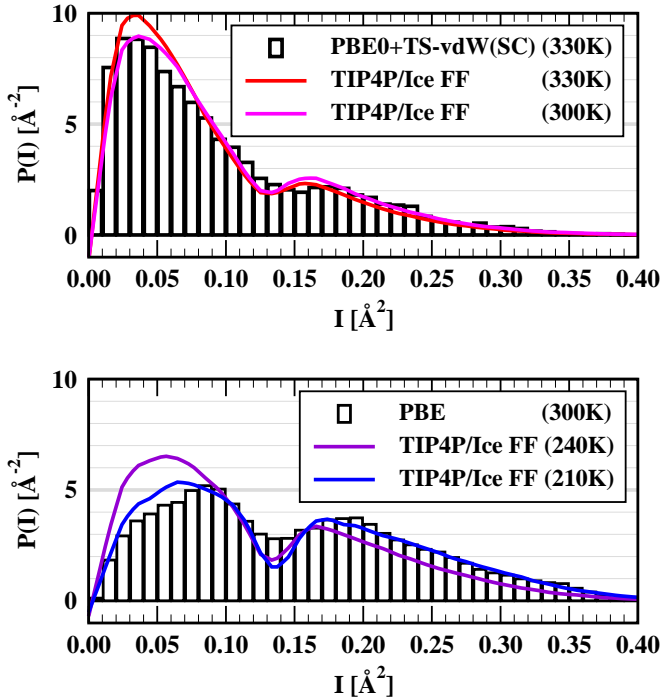}
\caption{\label{fig:plsi-ipes} Probability density distributions of the local structure index (P($I$)) in the inherent potential energy surfaces (IPES) of liquid water obtained from DFT-based AIMD simulations and classical force field (TIP4P/Ice) based simulations }
\end{center}
\end{figure}

A collection of local potential energy minima resulting from systematic quenches along a liquid trajectory, the IPES is a powerful concept that allows us to separate packing from thermal motion effects~\cite{StillingerJCP1984,Stillinger1984}.
To study the LSI distribution in the corresponding IPES of ambient liquid water, approximately 100 configuration snapshots were taken at 0.25 ps intervals from the PBE (300 K) and PBE0+TS-vdW(SC) (330 K) simulations and rapidly quenched to the nearest local potential energy minima~\textit{via} second-order damped dynamics~\cite{Tassone1994}.
Upon removal of the thermal energy from these ambient liquid water structures, the LSI distributions were found to be bimodal in shape (as opposed to the unimodal distributions found above in Section~\ref{subsec:AmbientLSI}) and are depicted in Figure~\ref{fig:plsi-ipes}.
For both of these XC functionals, the two peaks in the bimodal P($I$) distributions are separated by distinguishable minima ($I_{\rm min}$) located at approximately 0.14--0.16~\AA$^2$ (which should be contrasted against the fact that $I_{\rm min} =$ 0.13--0.14 ~\AA$^2$ for the TIP4P/Ice force field over the temperature range of 210--330 K).
Considering the fact that our IPES P($I$) have a resolution of $>$ 0.01~\AA$^2$ (due to the relatively low number of uncorrelated configurations accessed in our AIMD simulation), the position of this so-called isobestic point seems to be slightly dependent on these rather distinct \textit{ab initio} and classical potentials.
In this regard, previous studies of the IPES in liquid water based on classical force fields have reported that the position of this minimum is nearly invariant over a wide range of temperatures and pressures~\cite{Appignanesi2009,Accordino2011,Wikfeldt2011}.
However, we note in passing that the position of the isobestic point is also weakly dependent on the choice of the cutoff distance used in Equation~(\ref{eq:lsi}) to define the LSI distribution of radial distances~\cite{Accordino2011}.

In the IPES, we found that the relative heights of the two peaks in the bimodal P($I$) distribution obtained with the PBE XC functional are significantly different than those obtained at the PBE0+TS-vdW(SC) level of theory.
Based on the positions of $I_{\rm min}=0.14$ \AA$^2$~\cite{imin_footnote} in the IPES P($I$) distributions, the relative amount of high- and low-density-like molecules in liquid water generated by the PBE0+TS-vdW(SC) XC functional was found as 77\% and 23\%, respectively, whereas in the case of PBE they are almost equally distributed.
Hence, with thermal excitations removed, PBE0+TS-vdW(SC) liquid water is still predominantly comprised of water molecules in high-density-like locally disordered environments, while PBE liquid water is essentially an equal mixture of low- and high-LSI sites.
As observed above in Section~\ref{subsec:AmbientLSI} for the thermally excited P($I$) distributions, the IPES P($I$) obtained from PBE0+TS-vdW(SC) closely resembles the TIP4P/Ice classical force field results obtained at $\sim$ 300 K, whereas the PBE distribution is in closer agreement to the TIP4P/Ice data obtained at $\sim$ 210 K.
This finding is again suggestive that the IPES of liquid water simulated at ambient temperature with the PBE XC functional is more similar to the IPES of a deeply supercooled liquid.

This P($I$) analysis at the IPES again exemplifies the importance of exact exchange and vdW/dispersion forces in generating an accurate microscopic description of ambient liquid water.
The increased population of high-density-like molecular environments found in PBE0+TS-vdW(SC) liquid water at ambient conditions and at the corresponding IPES originates from the collective effects of weakening the directional hydrogen bonds (facilitated by a reduction in the self-interaction error) and strengthening the non-directional vdW/dispersion interactions.
Since this vdW-inclusive hybrid XC functional is able to describe ambient liquid water with good accuracy~\cite{Distasio2014}, we perform the next set of analyses at the PBE0+TS-vdW(SC) level of theory \textit{only}, in order to further our understanding of the spatial and temporal correlations among molecules identified by low- and high-density-like environments in liquid water in the IPES of ambient liquid water.

\subsection{Signature of Polyamorphism in the IPES of Ambient Liquid Water \label{subsec:SpaceCorrelation}}

\begin{figure}[t!]
\begin{center}
\fbox{\includegraphics[width=8.4cm]{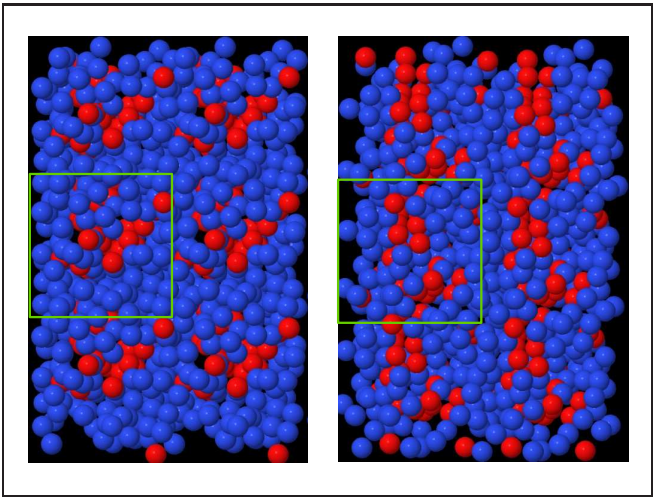}}
\caption{\label{fig:snapshot} Snapshots of typical liquid water structures (configurations) found in the IPES of ambient liquid water computed with the PBE0+TS-vdW(SC) XC potential at 330K. The blue spheres represent water molecules characterized by $I<I_{\rm min}$ (\textit{i.e.}, high-density-like water molecules), whereas the red spheres represent water molecules characterized by $I>I_{\rm min}$ (\textit{i.e.}, low-density-like water molecules). The green box refers to the simulation cell containing 128 water molecules with only the oxygen atoms shown above for clarity.}
\end{center}
\end{figure}

The bimodality observed in the IPES of liquid water allows us to characterize molecules having high- and low-density-like local structural environments based on the position of the isobestic point, $I_{\rm min}=0.14$ \AA$^2$~\cite{imin_footnote}, by dividing water molecules according to $I<I_{\rm min}$ and $I>I_{\rm min}$, respectively (see Section~\ref{subsec:InherentLSI}).
In the IPES of liquid water simulated with PBE0+TS-vdW(SC) at 330 K, the ratio between high- and low-density-like molecules is found to be approximately 3:1 based on this distinction.
When the effects of thermal motions are removed in the IPES, low-density-like molecules form small connected clusters amid the larger relative population of high-density-like molecules (see Figure~\ref{fig:snapshot}).
In our simulation cell, which contains 128 water molecules, we obtained 29$\pm$7 low-density-like molecules in the IPES on average and found cluster sizes among them that contained up to 34 molecules.
In this case, a molecule is considered to be part of a cluster if it is located within 3.33 \AA\ of any other water molecule present in the cluster (based on the oxygen-oxygen distance).
The clusters formed among the low-density-like water molecules do not span the length of the simulation cell. 
However, these small connected clusters may act as precursors for the growth of larger clusters of low-density-like molecules upon further cooling, a trend which was found in other theoretical studies~\cite{Appignanesi2009,Accordino2011,Wikfeldt2011}.

\begin{figure}[t!]
\begin{center}
\includegraphics[width=8.50cm]{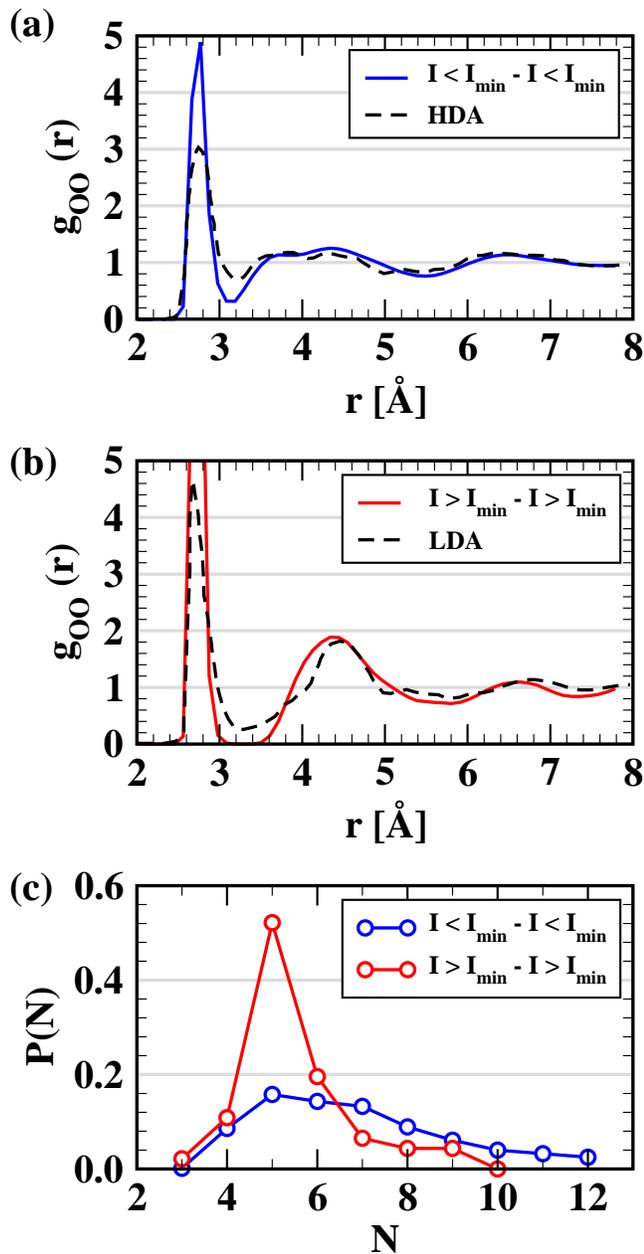}
\caption{\label{fig:gr} Comparison of the oxygen-oxygen radial distribution functions, $g_{\rm OO}(r)$, computed \textit{among} water molecules characterized by (a) $I<I_{\rm min}$ only and (b) $I>I_{\rm min}$ only based on liquid water structures (configurations) taken from the IPES generated with the PBE0+TS-vdW(SC) XC potential, with the experimental structures of the low-temperature (80 K) high-density (HDA) and low-density (LDA) amorphous ice phases, respectively. The experimental $g_{\rm OO}(r)$ were obtained using empirical potential structure refinement (EPSR) of neutron diffraction data~\cite{Finney2002}. Probability distributions of the hydrogen-bonded $N$-membered rings, P($N$), computed from structures on the PBE0+TS-vdW(SC) IPES are depicted in panel (c). All P($N$) were normalized to unity and therefore do not reflect the total number of rings of a given size.}
\end{center}
\end{figure}

In order to quantitatively study how molecules characterized by these distinct local environments are spatially correlated in the IPES, we have computed oxygen-oxygen radial distribution functions, $g_{\rm OO}(r)$, separately \textit{among} the high- and low-density-like water molecules.
In doing so, we found a remarkable similarity between the $g_{\rm OO}(r)$ computed \textit{among} the high-density-like (\textit {i.e.}, with $I<I_{\rm min}$) water molecules only and the low-density-like (\textit {i.e.}, with $I>I_{\rm min}$) water molecules only with the experimentally observed $g_{\rm OO}(r)$ for the high-density (HDA) and low-density (LDA) amorphous ice phases, respectively.
In Figure~\ref{fig:gr}(a)-(b), these theoretical $g_{\rm OO}(r)$ are compared against the $g_{\rm OO}(r)$ corresponding to the low-temperature (80 K) HDA and LDA phases, which were obtained from empirical potential structure refinement (EPSR) of experimental neutron diffraction data~\cite{Finney2002}.
From this picture, two different molecular environments are clearly evident---in the case of $I<I_{\rm min}$ (see Figure~\ref{fig:gr}(a)), the interstitial region between the first and second coordination shells is largely populated, whereas the separation between the first and second coordination shells is prominent in the case of $I>I_{\rm min}$ (see Figure~\ref{fig:gr}(b)).
Beyond the first coordination shell, \textit{i.e.}, in the intermediate range order (IRO) region, correspondence between the theoretical and experimental $g_{\rm OO}(r)$ is remarkable up to the length of the simulation cell employed herein.
The prominent feature of a much shallower and broader second shell present in HDA is observed among IPES molecules with $I<I_{\rm min}$ (see Figure~\ref{fig:gr}(a)), whereas a much sharper and narrower second shell at $\sim$ 4.5 \AA\ present in LDA is observed among IPES molecules with $I>I_{\rm min}$ (see Figure~\ref{fig:gr}(b)).
The correspondence between theoretical and experimental $g_{\rm OO}(r)$ in the region of the first coordination shell, which defines the short range order (SRO), is significantly effected by the thermal and quantal motions present in HDA and LDA.
The complete removal of thermal effects in the quenched structures obtained at the IPES is reflected in the fact that the first peaks of \textit{both} theoretical $g_{\rm OO}(r)$ are too sharp when compared to the experimental $g_{\rm OO}(r)$ of the amorphous structures, which show a much broader first peak.
Due to the very same reason, we also observed a decrease in the population of molecules in the interstitial regions of the theoretical $g_{\rm OO}(r)$ when compared to the analogous experimental quantities.
As such, these findings clearly suggest that the IPES of ambient liquid water contains the signatures of the experimentally observed polymorphism in the amorphous phases of ice.

Apart from the distinctions observed in the $g_{\rm OO}(r)$ above, the structural differences between HDA and LDA also lie in the topologies of their underlying hydrogen bond networks. 
These differential topologies can be quantified using ring statistics analysis---a tool which has been instrumental in theoretically characterizing and delineating the evolution of the HDA and LDA phases~\cite{Martonak2004,Martonak2005}.
In the case of LDA, which is more ordered, one finds a narrower distribution of rings centered about the most predominant six-membered ring, whereas in the case of HDA, which is more disordered, one finds a much broader distribution of rings with a substantial population having more than six members.
In fact, the existence of larger rings provides a pathway of having more compact and disordered packing of water molecules in the amorphous ice and liquid phases of water~\cite{Martonak2004,Martonak2005}.
Here, we have used a geometric definition of intact hydrogen bonds~\cite{hbond_footnote} in conjunction with the criteria of King~\cite{King1967} to find the shortest-path closed rings containing $N$ members \textit{among} water molecules characterized by $I>I_{\rm min}$ only and $I<I_{\rm min}$ only, and computed the resultant normalized probability distributions, P($N$), depicted in Figure~\ref{fig:gr}(c).
From this figure, it can readily be seen that the hydrogen bond networks connecting water molecules with the same LSI identities have very different topologies.
For instance, we found a broad distribution of rings and a tail that reaches out to $N=12$ for the high-density-like molecules.
Closed rings containing $N>12$ members were omitted because these rings start to include periodic images of the molecules forming the ring due to the finite size of the simulation cells employed in this work.
On the other hand, the P($N$) obtained from low-density-like molecules is much narrower and does not contain rings with more than $N=9$ members, consistent with the observation that low-density clusters do not span the length of the cell.
As such, these difference in the P($N$) of low- and high-density-like clusters bears strong similarities with those found for the LDA and HDA ice phases~\cite{Martonak2004,Martonak2005}.

Instead of a predominance of $N=6$ membered rings, as expected in the structures of LDA~\cite{Martonak2004,Martonak2005}, we found that the largest majority of rings contain $N=5$ members with significantly large populations of smaller rings among the high- and low-density-like water molecules in the IPES of ambient liquid water.
As such, this higher population of shorter rings (containing $N<6$ members) is reflective of the fact that the IPES of ambient liquid water does not contain true phases of either HDA or LDA.
In order to obtain the HDA and LDA phases of ice, structural relaxations on much larger length and time scales are required---relaxations which are certainly not achieved with instantaneous quenching to the nearest local potential energy minimum, the technique employed in this work to generate the corresponding IPES.
However, the topological differences in the underlying hydrogen bond networks observed here is striking and consistent with the presence of a first-order phase boundary between the amorphous phases of ice, as noted in Ref.~\cite{Martonak2004,Martonak2005}.
Furthermore, this finding complements the distinctive features of HDA and LDA observed \textit{via} the $g_{\rm OO}(r)$ above.

\subsection{Autocorrelation of the LSI Order Parameter in Ambient Liquid Water \label{subsec:TimeCorrelation}}

All of the analyses and discussion provided above in Section~\ref{subsec:SpaceCorrelation} were only made possible with the knowledge of the liquid water structures at the IPES, obtained by removing the thermal motion effects from ambient liquid water.
At ambient temperatures, these thermal motions smear the distinctions made based on $I_{\rm min}$ in the IPES and effectively result in the unimodal P($I$) discussed in Section~\ref{subsec:AmbientLSI} (see Figure~\ref{fig:plsi}).
Hence, these thermal motions introduce decay in the fluctuations in the local structural environment surrounding a given water molecule.

\begin{figure}[t!]
\begin{center}
\includegraphics[width=8.75cm]{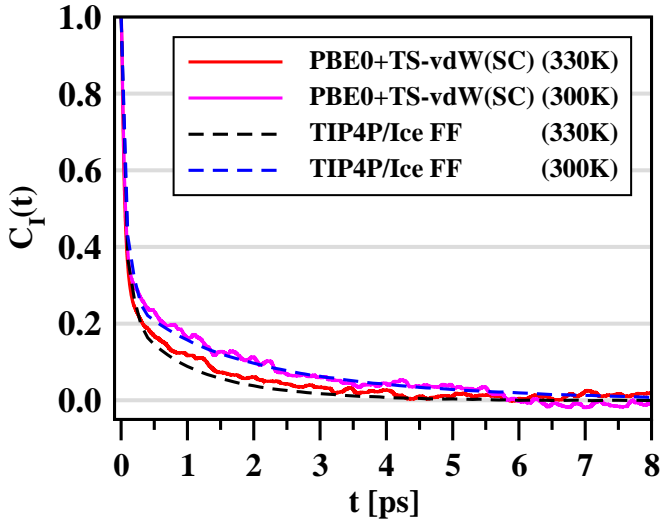}
\caption{\label{fig:acf} Autocorrelation function of the local structure index order parameter, $C_I(t)$, computed from DFT-based AIMD (PBE0+TS-vdW(SC)) and classical force field (TIP4P/Ice) based simulations of liquid water at temperatures in the range of 300--330 K.}
\end{center}
\end{figure}

To investigate the temporal evolution of these thermal fluctuations in the local structural environment, we have computed the autocorrelation function, $C_I(t)$, of the LSI order parameter, defined as:
\begin{equation}
\label{eq:acf}
C_I(t)=\frac{\langle \delta I(0) \delta I(t) \rangle}{\langle \delta I(0) \delta I(0) \rangle} \,,
\end{equation}
in which $\delta I(0)$ ($\delta I(t)$) is defined for a given water molecule as the difference between $I$ given by Equation~(\ref{eq:lsi}) at time $t=0$ ($t=t$) and the ensemble average $\overline I$ over all molecules and MD snapshots, \textit{i.e.}, $\delta I \equiv I - \overline I$.
As shown in Figure~\ref{fig:acf}, the LSI autocorrelation function decays to zero within approximately 4 ps for the case of PBE0+TS-vdW(SC) liquid water computed at 330 K.
Here we note that this $\sim$ 4 ps persistence time corresponding to different local molecular environments is similar to the time scales associated with the density-density fluctuations observed in large-scale classical simulations of ambient liquid water~\cite{English2011}.
For comparison, the $C_I(t)$ obtained from TIP4P/Ice liquid water at 330 K shows almost identical behavior with the AIMD results.
Upon reduction of the simulation temperature to 300 K, the autocorrelation functions for both PBE0+TS-vdW(SC) and TIP4P/Ice still remain extremely similar with correlation times increasing to $\sim$ 6 ps.

From Figure~\ref{fig:acf}, one can also notice that the decay rate of $C_I(t)$ below 0.5 ps occurs significantly faster and is essentially independent of the simulation temperature---an effect which is primarily due to the librational motion in liquid water.
Beyond 0.5 ps, the decay rate of $C_I(t)$ is much slower and is influenced by the collective thermal motions of the molecules---motions which are fairly sensitive to the simulation temperature.
These findings show that naturally occurring thermal fluctuations in ambient liquid water originate local LSI fluctuations which allow water molecules to sample different local structural environments.
Hence the presence of low- and high-density-like sites are rather transient, with a given molecule sampling the continuous distribution of local environments existing between these two distinct states over the course of approximately $\sim$ 4 ps.

In addition, we note in passing that the sluggish dynamics observed in liquid water simulated at 300 K utilizing the PBE XC functional are directly reflected in the persistence time between low- and high-density-like sites.
The $C_I(t)$ computed from PBE (300 K) liquid water decays at a much slower rate and vanishes at $\sim$ 14 ps, which is approximately 3 times longer than that computed from PBE0+TS-vdW(SC) (330 K) liquid water.

\section{Conclusions and Future Outlook \label{sec:Conclusions}}

In this work, we have performed a detailed computational study of the fluctuations in the local structural environments found in ambient liquid water using the local structure index (LSI) order parameter and a systematic hierarchy of DFT XC functionals.
For each of the functionals considered herein, we found probability distributions of the LSI order parameter, P($I$), that were all unimodal in shape and rapidly decaying from maxima located in the low-LSI range, indicating that a majority of the water molecules at ambient conditions are situated in locally disordered high-density-like environments.
We observed large variations among the relative populations of the continuous distribution of local structural environments existing in ambient liquid water as a function of the underlying XC potential---a trend that appears to be remarkably similar to the effects observed by changing the simulation temperature.
At the PBE level of theory, there exists a much higher percentage of molecules with high-LSI values, \textit{i.e.}, molecules which are characterized by low-density-like and more ordered local environments, and this finding provides additional microscopic structural insight into the overstructuring and sluggish nature of PBE liquid water simulated at ambient conditions.
Including the effects of exact exchange and vdW/dispersion interactions in the underlying XC potential resulted in a significant increase in the relative population of molecular sites with low-LSI values, which are more disordered and high-density like.
This increased population of high-density-like molecules originates from an energetic weakening of the hydrogen bonds in conjunction with an entropic stabilization of the resultant disordered water configurations, as facilitated by the inclusion of exact exchange and non-local vdW/dispersion forces.
As a result, ambient liquid water simulated with the vdW-inclusive hybrid functional, PBE0+TS-vdW(SC), is significantly less structured and more diffusive than PBE water with a substantial improvement in the agreement with experiment for a number of structural and electrostatic properties~\cite{Distasio2014}.

Focusing on liquid water generated at the PBE0+TS-vdW(SC) level of theory, we found the P($I$) to be unimodal at ambient temperatures, with removal of thermal excitation leading to a bimodal P($I$) distribution at the corresponding IPES.
This behavior is reflected in the density and spatial distribution of the low-density-like molecules.
In the IPES, the population of low-density-like molecules grows considerably and is accompanied by the formation of connected clusters which can include as many as 34 molecules in a (H$_2$O)$_{128}$ simulation cell.
These clusters do not span the length of the cell as found at liquid-liquid coexistence~\cite{Palmer2014}, but are sufficiently extended to allow for meaningful ring statistics analysis.
The presence of these low-density-like clusters amid the larger population of high-density-like molecules in the IPES of ambient water is noteworthy, as these clusters should act like seeds for the growth of larger low-density clusters upon cooling.
This behavior has indeed been observed in MD simulations with empirical force fields upon cooling~\cite{Appignanesi2009,Accordino2011,Wikfeldt2011} and is entirely consistent with the experimental observation of metastable $\mu$m-sized liquid droplets with low-density character~\cite{Sellberg2014}.
It is the presence of low-density and high-density fluctuations growing in size upon cooling that leads, under appropriate thermodynamic and kinetic conditions, to the formation of the LDA and/or HDA glasses and their liquid counterparts, LDL and/or HDL.
Our finding that the signatures of the intermediate range order (IRO) of LDA and HDA are both present in the IPES of ambient water is therefore quite remarkable.
It is also interesting that these signatures are associated to ring statistics distributions that correspond to distinct topologies of the underlying hydrogen bond network.
In this regard, a change of topology from low-density-like to high-density-like involves substantial breaking and reforming of hydrogen bonds accompanied by an associated energy barrier, which is consistent with the presence of a first-order transition boundary between LDA and HDA~\cite{Martonak2004,Martonak2005}.

Analysis of the autocorrelation of the LSI order parameter at ambient conditions shows that low- and high-density fluctuations are transient and decay with a lifetime of $\sim$ 4 ps at the PBE0+TS-vdW(SC) level of theory.
In other words, the naturally occurring thermal fluctuations allow a given water molecule to sample the continuous distribution of local structural environments between the two distinct high- and low-density states every $\sim$ 4 ps on average.
In this respect, it should also be noted that alternative two-state hydrogen bond models have often been invoked to explain the features observed in experiments probing the local structure of ambient water~\cite{Nilsson2011,Nilsson2012,Fuchs2008,Tokushima2008,Nilsson2010}.

As a final remark, we should mention that the quantitative details of the picture that we have presented in this work may change by further refining the \textit{ab initio} description of water with more accurate XC functionals that reduce the self-interaction error, as could possibly be achieved by fine-tuning the exchange correction~\cite{Skone2014}, and/or include a better description of vdW/dispersion interactions \textit{via} the inclusion of beyond-pairwise interactions, as provided, for example, by the recently proposed many-body dispersion (MBD) scheme~\cite{tkatchenko_prl_2012,distasio_pnas_2012,ambrosetti_jcp_2014,distasio_jpcm_2014}.
Further refinements are expected to result from a proper treatment of nuclear quantum effects \textit{via} explicit treatment of the quantum mechanical nature of the nuclei by the Feynman discretized path-integral approach.
However, all of these refinements are unlikely to change the qualitative picture exposed herein.

\begin{acknowledgments}
All authors acknowledge support from the Scientific Discovery through Advanced Computing (SciDAC) program through the Department of Energy (DOE) under Grant No. DE-SC0008626.
We also would like to acknowledge the DOE Computational Materials and Chemical Sciences Network (CMSCN) award DE-SC0005180, which provided support during the early stages of this work.
This research used resources of the National Energy Research Scientific Computing Center, which is supported by the Office of Science of the U.S. Department of Energy under Contract No. DE-AC02-05CH11231.
This research used resources of the Argonne Leadership Computing Facility at Argonne National Laboratory, which is supported by the Office of Science of the U.S. Department of Energy under contract DE-AC02-06CH11357.
Additional computational resources were provided by the Terascale Infrastructure for Groundbreaking Research in Science and Engineering (TIGRESS) High Performance Computing Center and Visualization Laboratory at Princeton University.
\end{acknowledgments}


\begin{thebibliography}{112}%
\makeatletter
\providecommand \@ifxundefined [1]{%
 \@ifx{#1\undefined}
}%
\providecommand \@ifnum [1]{%
 \ifnum #1\expandafter \@firstoftwo
 \else \expandafter \@secondoftwo
 \fi
}%
\providecommand \@ifx [1]{%
 \ifx #1\expandafter \@firstoftwo
 \else \expandafter \@secondoftwo
 \fi
}%
\providecommand \natexlab [1]{#1}%
\providecommand \enquote  [1]{``#1''}%
\providecommand \bibnamefont  [1]{#1}%
\providecommand \bibfnamefont [1]{#1}%
\providecommand \citenamefont [1]{#1}%
\providecommand \href@noop [0]{\@secondoftwo}%
\providecommand \href [0]{\begingroup \@sanitize@url \@href}%
\providecommand \@href[1]{\@@startlink{#1}\@@href}%
\providecommand \@@href[1]{\endgroup#1\@@endlink}%
\providecommand \@sanitize@url [0]{\catcode `\\12\catcode `\$12\catcode
  `\&12\catcode `\#12\catcode `\^12\catcode `\_12\catcode `\%12\relax}%
\providecommand \@@startlink[1]{}%
\providecommand \@@endlink[0]{}%
\providecommand \url  [0]{\begingroup\@sanitize@url \@url }%
\providecommand \@url [1]{\endgroup\@href {#1}{\urlprefix }}%
\providecommand \urlprefix  [0]{URL }%
\providecommand \Eprint [0]{\href }%
\providecommand \doibase [0]{http://dx.doi.org/}%
\providecommand \selectlanguage [0]{\@gobble}%
\providecommand \bibinfo  [0]{\@secondoftwo}%
\providecommand \bibfield  [0]{\@secondoftwo}%
\providecommand \translation [1]{[#1]}%
\providecommand \BibitemOpen [0]{}%
\providecommand \bibitemStop [0]{}%
\providecommand \bibitemNoStop [0]{.\EOS\space}%
\providecommand \EOS [0]{\spacefactor3000\relax}%
\providecommand \BibitemShut  [1]{\csname bibitem#1\endcsname}%
\let\auto@bib@innerbib\@empty
\bibitem [{\citenamefont {{DiStasio Jr.}}\ \emph {et~al.}(2014)\citenamefont
  {{DiStasio Jr.}}, \citenamefont {Santra}, \citenamefont {Li}, \citenamefont
  {Wu},\ and\ \citenamefont {Car}}]{Distasio2014}%
  \BibitemOpen
  \bibfield  {author} {\bibinfo {author} {\bibfnamefont {R.~A.}\ \bibnamefont
  {{DiStasio Jr.}}}, \bibinfo {author} {\bibfnamefont {B.}~\bibnamefont
  {Santra}}, \bibinfo {author} {\bibfnamefont {Z.}~\bibnamefont {Li}}, \bibinfo
  {author} {\bibfnamefont {X.}~\bibnamefont {Wu}}, \ and\ \bibinfo {author}
  {\bibfnamefont {R.}~\bibnamefont {Car}},\ }\href@noop {} {\bibfield
  {journal} {\bibinfo  {journal} {{\href{http://dx.doi.org/10.1063/1.4893377}
  {J. Chem. Phys.}}}\ }\textbf {\bibinfo {volume} {141}},\ \bibinfo {pages}
  {084502} (\bibinfo {year} {2014})}\BibitemShut {NoStop}%
\bibitem [{\citenamefont {Stillinger}(1980)}]{Stillinger1980}%
  \BibitemOpen
  \bibfield  {author} {\bibinfo {author} {\bibfnamefont {F.~H.}\ \bibnamefont
  {Stillinger}},\ }\href@noop {} {\bibfield  {journal} {\bibinfo  {journal}
  {\href{http://dx.doi.org/10.1126/science.209.4455.451}{Science}}\ }\textbf
  {\bibinfo {volume} {209}},\ \bibinfo {pages} {451} (\bibinfo {year}
  {1980})}\BibitemShut {NoStop}%
\bibitem [{\citenamefont {Franks}(2000)}]{franks_book_2000}%
  \BibitemOpen
  \bibfield  {author} {\bibinfo {author} {\bibfnamefont {F.}~\bibnamefont
  {Franks}},\ }\href@noop {} {\emph {\bibinfo {title} {Water: A Matrix of
  Life}}},\ \bibinfo {edition} {2nd}\ ed.\ (\bibinfo  {publisher} {The Royal
  Society of Chemistry},\ \bibinfo {address} {Cambridge},\ \bibinfo {year}
  {2000})\BibitemShut {NoStop}%
\bibitem [{\citenamefont {Santra}\ \emph {et~al.}(2011)\citenamefont {Santra},
  \citenamefont {Klime\v{s}}, \citenamefont {Alf\`{e}}, \citenamefont
  {Tkatchenko}, \citenamefont {Slater}, \citenamefont {Michaelides},
  \citenamefont {Car},\ and\ \citenamefont {Scheffler}}]{Santra2011}%
  \BibitemOpen
  \bibfield  {author} {\bibinfo {author} {\bibfnamefont {B.}~\bibnamefont
  {Santra}}, \bibinfo {author} {\bibfnamefont {J.}~\bibnamefont {Klime\v{s}}},
  \bibinfo {author} {\bibfnamefont {D.}~\bibnamefont {Alf\`{e}}}, \bibinfo
  {author} {\bibfnamefont {A.}~\bibnamefont {Tkatchenko}}, \bibinfo {author}
  {\bibfnamefont {B.}~\bibnamefont {Slater}}, \bibinfo {author} {\bibfnamefont
  {A.}~\bibnamefont {Michaelides}}, \bibinfo {author} {\bibfnamefont
  {R.}~\bibnamefont {Car}}, \ and\ \bibinfo {author} {\bibfnamefont
  {M.}~\bibnamefont {Scheffler}},\ }\href@noop {} {\bibfield  {journal}
  {\bibinfo  {journal}
  {\href{http://dx.doi.org/10.1103/PhysRevLett.107.185701}{Phys. Rev. Lett.}}\
  }\textbf {\bibinfo {volume} {107}},\ \bibinfo {pages} {185701} (\bibinfo
  {year} {2011})}\BibitemShut {NoStop}%
\bibitem [{\citenamefont {Santra}\ \emph {et~al.}(2013)\citenamefont {Santra},
  \citenamefont {Klime\v{s}}, \citenamefont {Tkatchenko}, \citenamefont
  {Alf\`e}, \citenamefont {Slater}, \citenamefont {Michaelides}, \citenamefont
  {Car},\ and\ \citenamefont {Scheffler}}]{Santra2013}%
  \BibitemOpen
  \bibfield  {author} {\bibinfo {author} {\bibfnamefont {B.}~\bibnamefont
  {Santra}}, \bibinfo {author} {\bibfnamefont {J.}~\bibnamefont {Klime\v{s}}},
  \bibinfo {author} {\bibfnamefont {A.}~\bibnamefont {Tkatchenko}}, \bibinfo
  {author} {\bibfnamefont {D.}~\bibnamefont {Alf\`e}}, \bibinfo {author}
  {\bibfnamefont {B.}~\bibnamefont {Slater}}, \bibinfo {author} {\bibfnamefont
  {A.}~\bibnamefont {Michaelides}}, \bibinfo {author} {\bibfnamefont
  {R.}~\bibnamefont {Car}}, \ and\ \bibinfo {author} {\bibfnamefont
  {M.}~\bibnamefont {Scheffler}},\ }\href@noop {} {\bibfield  {journal}
  {\bibinfo  {journal} {\href{http://dx.doi.org/10.1063/1.3012573}{J. Chem.
  Phys.}}\ }\textbf {\bibinfo {volume} {139}},\ \bibinfo {pages} {154702}
  (\bibinfo {year} {2013})}\BibitemShut {NoStop}%
\bibitem [{\citenamefont {Santra}\ \emph {et~al.}(2009)\citenamefont {Santra},
  \citenamefont {Michaelides},\ and\ \citenamefont {Scheffler}}]{Santra2009}%
  \BibitemOpen
  \bibfield  {author} {\bibinfo {author} {\bibfnamefont {B.}~\bibnamefont
  {Santra}}, \bibinfo {author} {\bibfnamefont {A.}~\bibnamefont {Michaelides}},
  \ and\ \bibinfo {author} {\bibfnamefont {M.}~\bibnamefont {Scheffler}},\
  }\href@noop {} {\bibfield  {journal} {\bibinfo  {journal}
  {\href{http://dx.doi.org/10.1063/1.3236840}{J. Chem. Phys.}}\ }\textbf
  {\bibinfo {volume} {131}},\ \bibinfo {pages} {124509} (\bibinfo {year}
  {2009})}\BibitemShut {NoStop}%
\bibitem [{\citenamefont {Santra}\ \emph {et~al.}(2008)\citenamefont {Santra},
  \citenamefont {Michaelides}, \citenamefont {Fuchs}, \citenamefont
  {Tkatchenko}, \citenamefont {Filippi},\ and\ \citenamefont
  {Scheffler}}]{Santra2008}%
  \BibitemOpen
  \bibfield  {author} {\bibinfo {author} {\bibfnamefont {B.}~\bibnamefont
  {Santra}}, \bibinfo {author} {\bibfnamefont {A.}~\bibnamefont {Michaelides}},
  \bibinfo {author} {\bibfnamefont {M.}~\bibnamefont {Fuchs}}, \bibinfo
  {author} {\bibfnamefont {A.}~\bibnamefont {Tkatchenko}}, \bibinfo {author}
  {\bibfnamefont {C.}~\bibnamefont {Filippi}}, \ and\ \bibinfo {author}
  {\bibfnamefont {M.}~\bibnamefont {Scheffler}},\ }\href@noop {} {\bibfield
  {journal} {\bibinfo  {journal} {\href{http://dx.doi.org/10.1063/1.3012573}{J.
  Chem. Phys.}}\ }\textbf {\bibinfo {volume} {129}},\ \bibinfo {pages} {194111}
  (\bibinfo {year} {2008})}\BibitemShut {NoStop}%
\bibitem [{Ang()}]{Angell1982}%
  \BibitemOpen
  \href@noop {} {}\bibinfo {note} {{C. A. Angell (1982) in Water: A
  Comprehensive Treatise, ed F. Franks (Plenum, New York), Vol 7, pp
  1–81.}}\BibitemShut {Stop}%
\bibitem [{\citenamefont {Mishima}\ and\ \citenamefont
  {Stanley}(1998)}]{Mishima1998}%
  \BibitemOpen
  \bibfield  {author} {\bibinfo {author} {\bibfnamefont {O.}~\bibnamefont
  {Mishima}}\ and\ \bibinfo {author} {\bibfnamefont {H.~E.}\ \bibnamefont
  {Stanley}},\ }\href@noop {} {\bibfield  {journal} {\bibinfo  {journal}
  {\href{http://dx.doi.org/10.1038/24540}{Nature}}\ }\textbf {\bibinfo {volume}
  {396}},\ \bibinfo {pages} {329} (\bibinfo {year} {1998})}\BibitemShut
  {NoStop}%
\bibitem [{\citenamefont {Debenedetti}(2003)}]{Debenedetti2003}%
  \BibitemOpen
  \bibfield  {author} {\bibinfo {author} {\bibfnamefont {P.~G.}\ \bibnamefont
  {Debenedetti}},\ }\href@noop {} {\bibfield  {journal} {\bibinfo  {journal}
  {\href{http://dx.doi.org/10.1088/0953-8984/15/45/R01}{J. Phys.: Condens.
  Matter}}\ }\textbf {\bibinfo {volume} {15}},\ \bibinfo {pages} {R1669}
  (\bibinfo {year} {2003})}\BibitemShut {NoStop}%
\bibitem [{\citenamefont {Mishima}\ \emph {et~al.}(1985)\citenamefont
  {Mishima}, \citenamefont {Calvert},\ and\ \citenamefont
  {Whalley}}]{Mishima1985}%
  \BibitemOpen
  \bibfield  {author} {\bibinfo {author} {\bibfnamefont {O.}~\bibnamefont
  {Mishima}}, \bibinfo {author} {\bibfnamefont {L.~D.}\ \bibnamefont
  {Calvert}}, \ and\ \bibinfo {author} {\bibfnamefont {E.}~\bibnamefont
  {Whalley}},\ }\href@noop {} {\bibfield  {journal} {\bibinfo  {journal}
  {\href{http://dx.doi.org/10.1038/314076a0}{Nature}}\ }\textbf {\bibinfo
  {volume} {314}},\ \bibinfo {pages} {76} (\bibinfo {year} {1985})}\BibitemShut
  {NoStop}%
\bibitem [{\citenamefont {Mishima}\ \emph {et~al.}(1991)\citenamefont
  {Mishima}, \citenamefont {Takemura},\ and\ \citenamefont
  {Aoki}}]{Mishima1991}%
  \BibitemOpen
  \bibfield  {author} {\bibinfo {author} {\bibfnamefont {O.}~\bibnamefont
  {Mishima}}, \bibinfo {author} {\bibfnamefont {K.}~\bibnamefont {Takemura}}, \
  and\ \bibinfo {author} {\bibfnamefont {K.}~\bibnamefont {Aoki}},\ }\href@noop
  {} {\bibfield  {journal} {\bibinfo  {journal}
  {\href{http://dx.doi.org/10.1126/science.254.5030.406}{Science}}\ }\textbf
  {\bibinfo {volume} {254}},\ \bibinfo {pages} {406} (\bibinfo {year}
  {1991})}\BibitemShut {NoStop}%
\bibitem [{\citenamefont {Mishima}\ and\ \citenamefont
  {Suzuki}(2002)}]{Mishima2002}%
  \BibitemOpen
  \bibfield  {author} {\bibinfo {author} {\bibfnamefont {O.}~\bibnamefont
  {Mishima}}\ and\ \bibinfo {author} {\bibfnamefont {Y.}~\bibnamefont
  {Suzuki}},\ }\href@noop {} {\bibfield  {journal} {\bibinfo  {journal}
  {\href{http://dx.doi.org/10.1038/nature01106}{Nature}}\ }\textbf {\bibinfo
  {volume} {419}},\ \bibinfo {pages} {599} (\bibinfo {year}
  {2002})}\BibitemShut {NoStop}%
\bibitem [{\citenamefont {Mishima}(1994)}]{Mishima1994}%
  \BibitemOpen
  \bibfield  {author} {\bibinfo {author} {\bibfnamefont {O.}~\bibnamefont
  {Mishima}},\ }\href@noop {} {\bibfield  {journal} {\bibinfo  {journal}
  {\href{http://dx.doi.org/10.1063/1.467103}{J. Chem. Phys.}}\ }\textbf
  {\bibinfo {volume} {100}},\ \bibinfo {pages} {5910} (\bibinfo {year}
  {1994})}\BibitemShut {NoStop}%
\bibitem [{\citenamefont {Winkel}\ \emph {et~al.}(2008)\citenamefont {Winkel},
  \citenamefont {Elsaesser}, \citenamefont {Mayer},\ and\ \citenamefont
  {Loerting}}]{Winkel2008}%
  \BibitemOpen
  \bibfield  {author} {\bibinfo {author} {\bibfnamefont {K.}~\bibnamefont
  {Winkel}}, \bibinfo {author} {\bibfnamefont {M.~S.}\ \bibnamefont
  {Elsaesser}}, \bibinfo {author} {\bibfnamefont {E.}~\bibnamefont {Mayer}}, \
  and\ \bibinfo {author} {\bibfnamefont {T.}~\bibnamefont {Loerting}},\
  }\href@noop {} {\bibfield  {journal} {\bibinfo  {journal}
  {\href{http://dx.doi.org/10.1063/1.2830029}{J. Chem. Phys.}}\ }\textbf
  {\bibinfo {volume} {128}} (\bibinfo {year} {2008})}\BibitemShut {NoStop}%
\bibitem [{\citenamefont {Winkel}\ \emph {et~al.}(2011)\citenamefont {Winkel},
  \citenamefont {Mayer},\ and\ \citenamefont {Loerting}}]{Winkel2011}%
  \BibitemOpen
  \bibfield  {author} {\bibinfo {author} {\bibfnamefont {K.}~\bibnamefont
  {Winkel}}, \bibinfo {author} {\bibfnamefont {E.}~\bibnamefont {Mayer}}, \
  and\ \bibinfo {author} {\bibfnamefont {T.}~\bibnamefont {Loerting}},\
  }\href@noop {} {\bibfield  {journal} {\bibinfo  {journal}
  {\href{http://dx.doi.org/10.1038/10.1021/jp203985w}{J. Phys. Chem. B}}\
  }\textbf {\bibinfo {volume} {115}},\ \bibinfo {pages} {14141} (\bibinfo
  {year} {2011})}\BibitemShut {NoStop}%
\bibitem [{\citenamefont {Finney}\ \emph {et~al.}(2002)\citenamefont {Finney},
  \citenamefont {Hallbrucker}, \citenamefont {Kohl}, \citenamefont {Soper},\
  and\ \citenamefont {Bowron}}]{Finney2002}%
  \BibitemOpen
  \bibfield  {author} {\bibinfo {author} {\bibfnamefont {J.~L.}\ \bibnamefont
  {Finney}}, \bibinfo {author} {\bibfnamefont {A.}~\bibnamefont {Hallbrucker}},
  \bibinfo {author} {\bibfnamefont {I.}~\bibnamefont {Kohl}}, \bibinfo {author}
  {\bibfnamefont {A.~K.}\ \bibnamefont {Soper}}, \ and\ \bibinfo {author}
  {\bibfnamefont {D.~T.}\ \bibnamefont {Bowron}},\ }\href@noop {} {\bibfield
  {journal} {\bibinfo  {journal}
  {\href{http://dx.doi.org/10.1103/PhysRevLett.88.225503}{Phys. Rev. Lett.}}\
  }\textbf {\bibinfo {volume} {88}},\ \bibinfo {pages} {225503} (\bibinfo
  {year} {2002})}\BibitemShut {NoStop}%
\bibitem [{\citenamefont {Amann-Winkel}\ \emph {et~al.}(2013)\citenamefont
  {Amann-Winkel}, \citenamefont {Gainaru}, \citenamefont {Handle},
  \citenamefont {Seidl}, \citenamefont {Nelson}, \citenamefont {Böhmer},\ and\
  \citenamefont {Loerting}}]{Amann-Winkel2013}%
  \BibitemOpen
  \bibfield  {author} {\bibinfo {author} {\bibfnamefont {K.}~\bibnamefont
  {Amann-Winkel}}, \bibinfo {author} {\bibfnamefont {C.}~\bibnamefont
  {Gainaru}}, \bibinfo {author} {\bibfnamefont {P.~H.}\ \bibnamefont {Handle}},
  \bibinfo {author} {\bibfnamefont {M.}~\bibnamefont {Seidl}}, \bibinfo
  {author} {\bibfnamefont {H.}~\bibnamefont {Nelson}}, \bibinfo {author}
  {\bibfnamefont {R.}~\bibnamefont {Böhmer}}, \ and\ \bibinfo {author}
  {\bibfnamefont {T.}~\bibnamefont {Loerting}},\ }\href@noop {} {\bibfield
  {journal} {\bibinfo  {journal}
  {\href{http://dx.doi.org/10.1073/pnas.1311718110}{Proc. Natl. Acad. Sci.
  USA}}\ }\textbf {\bibinfo {volume} {110}},\ \bibinfo {pages} {17720}
  (\bibinfo {year} {2013})}\BibitemShut {NoStop}%
\bibitem [{\citenamefont {Andersson}\ and\ \citenamefont
  {Inaba}(2006)}]{Andersson2006}%
  \BibitemOpen
  \bibfield  {author} {\bibinfo {author} {\bibfnamefont {O.}~\bibnamefont
  {Andersson}}\ and\ \bibinfo {author} {\bibfnamefont {A.}~\bibnamefont
  {Inaba}},\ }\href@noop {} {\bibfield  {journal} {\bibinfo  {journal}
  {\href{http://dx.doi.org/10.1103/PhysRevB.74.184201}{Phys. Rev. B}}\ }\textbf
  {\bibinfo {volume} {74}},\ \bibinfo {pages} {184201} (\bibinfo {year}
  {2006})}\BibitemShut {NoStop}%
\bibitem [{\citenamefont {Andersson}(2011)}]{Andersson2011}%
  \BibitemOpen
  \bibfield  {author} {\bibinfo {author} {\bibfnamefont {O.}~\bibnamefont
  {Andersson}},\ }\href@noop {} {\bibfield  {journal} {\bibinfo  {journal}
  {\href{http://dx.doi.org/10.1073/pnas.1016520108}{Proc. Natl. Acad. Sci.
  USA}}\ }\textbf {\bibinfo {volume} {108}},\ \bibinfo {pages} {11013}
  (\bibinfo {year} {2011})}\BibitemShut {NoStop}%
\bibitem [{\citenamefont {Poole}\ \emph {et~al.}(1992)\citenamefont {Poole},
  \citenamefont {Sciortino}, \citenamefont {Essmann},\ and\ \citenamefont
  {Stanley}}]{Poole1992}%
  \BibitemOpen
  \bibfield  {author} {\bibinfo {author} {\bibfnamefont {P.~H.}\ \bibnamefont
  {Poole}}, \bibinfo {author} {\bibfnamefont {F.}~\bibnamefont {Sciortino}},
  \bibinfo {author} {\bibfnamefont {U.}~\bibnamefont {Essmann}}, \ and\
  \bibinfo {author} {\bibfnamefont {H.~E.}\ \bibnamefont {Stanley}},\
  }\href@noop {} {\bibfield  {journal} {\bibinfo  {journal}
  {\href{http://dx.doi.org/10.1038/360324a0}{Nature}}\ }\textbf {\bibinfo
  {volume} {360}},\ \bibinfo {pages} {324} (\bibinfo {year}
  {1992})}\BibitemShut {NoStop}%
\bibitem [{\citenamefont {Sellberg}\ \emph {et~al.}(2014)\citenamefont
  {Sellberg}, \citenamefont {Huang}, \citenamefont {McQueen}, \citenamefont
  {Loh}, \citenamefont {Laksmono}, \citenamefont {Schlesinger}, \citenamefont
  {Sierra}, \citenamefont {Nordlund}, \citenamefont {Hampton}, \citenamefont
  {Starodub}, \citenamefont {DePonte}, \citenamefont {Beye}, \citenamefont
  {Chen}, \citenamefont {Martin}, \citenamefont {Barty}, \citenamefont
  {Wikfeldt}, \citenamefont {Weiss}, \citenamefont {Caronna}, \citenamefont
  {Feldkamp}, \citenamefont {Skinner}, \citenamefont {Seibert}, \citenamefont
  {Messerschmidt}, \citenamefont {Williams}, \citenamefont {Boutet},
  \citenamefont {Pettersson}, \citenamefont {Bogan},\ and\ \citenamefont
  {Nilsson}}]{Sellberg2014}%
  \BibitemOpen
  \bibfield  {author} {\bibinfo {author} {\bibfnamefont {J.~A.}\ \bibnamefont
  {Sellberg}}, \bibinfo {author} {\bibfnamefont {C.}~\bibnamefont {Huang}},
  \bibinfo {author} {\bibfnamefont {T.~A.}\ \bibnamefont {McQueen}}, \bibinfo
  {author} {\bibfnamefont {N.~D.}\ \bibnamefont {Loh}}, \bibinfo {author}
  {\bibfnamefont {H.}~\bibnamefont {Laksmono}}, \bibinfo {author}
  {\bibfnamefont {D.}~\bibnamefont {Schlesinger}}, \bibinfo {author}
  {\bibfnamefont {R.~G.}\ \bibnamefont {Sierra}}, \bibinfo {author}
  {\bibfnamefont {D.}~\bibnamefont {Nordlund}}, \bibinfo {author}
  {\bibfnamefont {C.~Y.}\ \bibnamefont {Hampton}}, \bibinfo {author}
  {\bibfnamefont {D.}~\bibnamefont {Starodub}}, \bibinfo {author}
  {\bibfnamefont {D.~P.}\ \bibnamefont {DePonte}}, \bibinfo {author}
  {\bibfnamefont {M.}~\bibnamefont {Beye}}, \bibinfo {author} {\bibfnamefont
  {C.}~\bibnamefont {Chen}}, \bibinfo {author} {\bibfnamefont {A.~V.}\
  \bibnamefont {Martin}}, \bibinfo {author} {\bibfnamefont {A.}~\bibnamefont
  {Barty}}, \bibinfo {author} {\bibfnamefont {K.~T.}\ \bibnamefont {Wikfeldt}},
  \bibinfo {author} {\bibfnamefont {T.~M.}\ \bibnamefont {Weiss}}, \bibinfo
  {author} {\bibfnamefont {C.}~\bibnamefont {Caronna}}, \bibinfo {author}
  {\bibfnamefont {J.}~\bibnamefont {Feldkamp}}, \bibinfo {author}
  {\bibfnamefont {L.~B.}\ \bibnamefont {Skinner}}, \bibinfo {author}
  {\bibfnamefont {M.~M.}\ \bibnamefont {Seibert}}, \bibinfo {author}
  {\bibfnamefont {M.}~\bibnamefont {Messerschmidt}}, \bibinfo {author}
  {\bibfnamefont {G.~J.}\ \bibnamefont {Williams}}, \bibinfo {author}
  {\bibfnamefont {S.}~\bibnamefont {Boutet}}, \bibinfo {author} {\bibfnamefont
  {L.~G.~M.}\ \bibnamefont {Pettersson}}, \bibinfo {author} {\bibfnamefont
  {M.~J.}\ \bibnamefont {Bogan}}, \ and\ \bibinfo {author} {\bibfnamefont
  {A.}~\bibnamefont {Nilsson}},\ }\href@noop {} {\bibfield  {journal} {\bibinfo
   {journal} {\href{http://dx.doi.org/10.1038/nature13266}{Nature}}\ }\textbf
  {\bibinfo {volume} {510}},\ \bibinfo {pages} {381} (\bibinfo {year}
  {2014})}\BibitemShut {NoStop}%
\bibitem [{\citenamefont {Palmer}\ \emph {et~al.}(2014)\citenamefont {Palmer},
  \citenamefont {Martelli}, \citenamefont {Liu}, \citenamefont {Car},
  \citenamefont {Panagiotopoulos},\ and\ \citenamefont
  {Debenedetti}}]{Palmer2014}%
  \BibitemOpen
  \bibfield  {author} {\bibinfo {author} {\bibfnamefont {J.~C.}\ \bibnamefont
  {Palmer}}, \bibinfo {author} {\bibfnamefont {F.}~\bibnamefont {Martelli}},
  \bibinfo {author} {\bibfnamefont {Y.}~\bibnamefont {Liu}}, \bibinfo {author}
  {\bibfnamefont {R.}~\bibnamefont {Car}}, \bibinfo {author} {\bibfnamefont
  {A.~Z.}\ \bibnamefont {Panagiotopoulos}}, \ and\ \bibinfo {author}
  {\bibfnamefont {P.~G.}\ \bibnamefont {Debenedetti}},\ }\href@noop {}
  {\bibfield  {journal} {\bibinfo  {journal}
  {\href{http://dx.doi.org/10.1038/nature13405}{Nature}}\ }\textbf {\bibinfo
  {volume} {509}},\ \bibinfo {pages} {385} (\bibinfo {year}
  {2014})}\BibitemShut {NoStop}%
\bibitem [{\citenamefont {Tanaka}(2002)}]{Tanaka2002}%
  \BibitemOpen
  \bibfield  {author} {\bibinfo {author} {\bibfnamefont {H.}~\bibnamefont
  {Tanaka}},\ }\href@noop {} {\bibfield  {journal} {\bibinfo  {journal}
  {\href{http://dx.doi.org/10.1103/PhysRevB.66.064202}{Phys. Rev. B}}\ }\textbf
  {\bibinfo {volume} {66}},\ \bibinfo {pages} {064202} (\bibinfo {year}
  {2002})}\BibitemShut {NoStop}%
\bibitem [{\citenamefont {Stillinger}\ and\ \citenamefont
  {Rahman}(1974)}]{ST2model}%
  \BibitemOpen
  \bibfield  {author} {\bibinfo {author} {\bibfnamefont {F.~H.}\ \bibnamefont
  {Stillinger}}\ and\ \bibinfo {author} {\bibfnamefont {A.}~\bibnamefont
  {Rahman}},\ }\href@noop {} {\bibfield  {journal} {\bibinfo  {journal}
  {\href{http://dx.doi.org/10.1063/1.1681229}{J. Chem. Phys.}}\ }\textbf
  {\bibinfo {volume} {60}},\ \bibinfo {pages} {1545} (\bibinfo {year}
  {1974})}\BibitemShut {NoStop}%
\bibitem [{\citenamefont {Moore}\ and\ \citenamefont
  {Molinero}(2010)}]{Moore2010}%
  \BibitemOpen
  \bibfield  {author} {\bibinfo {author} {\bibfnamefont {E.~B.}\ \bibnamefont
  {Moore}}\ and\ \bibinfo {author} {\bibfnamefont {V.}~\bibnamefont
  {Molinero}},\ }\href@noop {} {\bibfield  {journal} {\bibinfo  {journal}
  {\href{http://dx.doi.org/10.1063/1.3451112}{J. Chem. Phys.}}\ }\textbf
  {\bibinfo {volume} {132}},\ \bibinfo {pages} {244504} (\bibinfo {year}
  {2010})}\BibitemShut {NoStop}%
\bibitem [{\citenamefont {Soper}\ and\ \citenamefont
  {Ricci}(2000)}]{Soper2000}%
  \BibitemOpen
  \bibfield  {author} {\bibinfo {author} {\bibfnamefont {A.~K.}\ \bibnamefont
  {Soper}}\ and\ \bibinfo {author} {\bibfnamefont {M.~A.}\ \bibnamefont
  {Ricci}},\ }\href@noop {} {\bibfield  {journal} {\bibinfo  {journal}
  {\href{http://dx.doi.org/10.1103/PhysRevLett.84.2881}{Phys. Rev. Lett.}}\
  }\textbf {\bibinfo {volume} {84}},\ \bibinfo {pages} {2881} (\bibinfo {year}
  {2000})}\BibitemShut {NoStop}%
\bibitem [{\citenamefont {Huang}\ \emph {et~al.}(2009)\citenamefont {Huang},
  \citenamefont {Wikfeldt}, \citenamefont {Tokushima}, \citenamefont
  {Nordlund}, \citenamefont {Harada}, \citenamefont {Bergmann}, \citenamefont
  {Niebuhr}, \citenamefont {Weiss}, \citenamefont {Horikawa}, \citenamefont
  {Leetmaa}, \citenamefont {Ljungberg}, \citenamefont {Takahashi},
  \citenamefont {Lenz}, \citenamefont {Ojamäe}, \citenamefont {Lyubartsev},
  \citenamefont {Shin}, \citenamefont {Pettersson},\ and\ \citenamefont
  {Nilsson}}]{Huang2009}%
  \BibitemOpen
  \bibfield  {author} {\bibinfo {author} {\bibfnamefont {C.}~\bibnamefont
  {Huang}}, \bibinfo {author} {\bibfnamefont {K.~T.}\ \bibnamefont {Wikfeldt}},
  \bibinfo {author} {\bibfnamefont {T.}~\bibnamefont {Tokushima}}, \bibinfo
  {author} {\bibfnamefont {D.}~\bibnamefont {Nordlund}}, \bibinfo {author}
  {\bibfnamefont {Y.}~\bibnamefont {Harada}}, \bibinfo {author} {\bibfnamefont
  {U.}~\bibnamefont {Bergmann}}, \bibinfo {author} {\bibfnamefont
  {M.}~\bibnamefont {Niebuhr}}, \bibinfo {author} {\bibfnamefont {T.~M.}\
  \bibnamefont {Weiss}}, \bibinfo {author} {\bibfnamefont {Y.}~\bibnamefont
  {Horikawa}}, \bibinfo {author} {\bibfnamefont {M.}~\bibnamefont {Leetmaa}},
  \bibinfo {author} {\bibfnamefont {M.~P.}\ \bibnamefont {Ljungberg}}, \bibinfo
  {author} {\bibfnamefont {O.}~\bibnamefont {Takahashi}}, \bibinfo {author}
  {\bibfnamefont {A.}~\bibnamefont {Lenz}}, \bibinfo {author} {\bibfnamefont
  {L.}~\bibnamefont {Ojamäe}}, \bibinfo {author} {\bibfnamefont {A.~P.}\
  \bibnamefont {Lyubartsev}}, \bibinfo {author} {\bibfnamefont
  {S.}~\bibnamefont {Shin}}, \bibinfo {author} {\bibfnamefont {L.~G.~M.}\
  \bibnamefont {Pettersson}}, \ and\ \bibinfo {author} {\bibfnamefont
  {A.}~\bibnamefont {Nilsson}},\ }\href@noop {} {\bibfield  {journal} {\bibinfo
   {journal} {\href{http://dx.doi.org/10.1073/pnas.0904743106}{Proc. Natl.
  Acad. Sci. USA}}\ }\textbf {\bibinfo {volume} {106}},\ \bibinfo {pages}
  {15214} (\bibinfo {year} {2009})}\BibitemShut {NoStop}%
\bibitem [{\citenamefont {Nilsson}\ and\ \citenamefont
  {Pettersson}(2011)}]{Nilsson2011}%
  \BibitemOpen
  \bibfield  {author} {\bibinfo {author} {\bibfnamefont {A.}~\bibnamefont
  {Nilsson}}\ and\ \bibinfo {author} {\bibfnamefont {L.}~\bibnamefont
  {Pettersson}},\ }\href@noop {} {\bibfield  {journal} {\bibinfo  {journal}
  {\href{http://dx.doi.org/10.1016/j.chemphys.2011.07.021}{Chem. Phys.}}\
  }\textbf {\bibinfo {volume} {389}},\ \bibinfo {pages} {1} (\bibinfo {year}
  {2011})}\BibitemShut {NoStop}%
\bibitem [{\citenamefont {Nilsson}\ \emph {et~al.}(2012)\citenamefont
  {Nilsson}, \citenamefont {Huang},\ and\ \citenamefont
  {Pettersson}}]{Nilsson2012}%
  \BibitemOpen
  \bibfield  {author} {\bibinfo {author} {\bibfnamefont {A.}~\bibnamefont
  {Nilsson}}, \bibinfo {author} {\bibfnamefont {C.}~\bibnamefont {Huang}}, \
  and\ \bibinfo {author} {\bibfnamefont {L.~G.}\ \bibnamefont {Pettersson}},\
  }\href@noop {} {\bibfield  {journal} {\bibinfo  {journal}
  {\href{http://dx.doi.org/10.1016/j.molliq.2012.06.021}{J. Mol. Liq.}}\
  }\textbf {\bibinfo {volume} {176}},\ \bibinfo {pages} {2} (\bibinfo {year}
  {2012})}\BibitemShut {NoStop}%
\bibitem [{\citenamefont {Shiratani}\ and\ \citenamefont
  {Sasai}(1996)}]{Shiratani1996}%
  \BibitemOpen
  \bibfield  {author} {\bibinfo {author} {\bibfnamefont {E.}~\bibnamefont
  {Shiratani}}\ and\ \bibinfo {author} {\bibfnamefont {M.}~\bibnamefont
  {Sasai}},\ }\href@noop {} {\bibfield  {journal} {\bibinfo  {journal}
  {\href{http://dx.doi.org/10.1063/1.471475}{J. Chem. Phys.}}\ }\textbf
  {\bibinfo {volume} {104}},\ \bibinfo {pages} {7671} (\bibinfo {year}
  {1996})}\BibitemShut {NoStop}%
\bibitem [{\citenamefont {Shiratani}\ and\ \citenamefont
  {Sasai}(1998)}]{Shiratani1998}%
  \BibitemOpen
  \bibfield  {author} {\bibinfo {author} {\bibfnamefont {E.}~\bibnamefont
  {Shiratani}}\ and\ \bibinfo {author} {\bibfnamefont {M.}~\bibnamefont
  {Sasai}},\ }\href@noop {} {\bibfield  {journal} {\bibinfo  {journal}
  {\href{http://dx.doi.org/10.1063/1.475723}{J. Chem. Phys.}}\ }\textbf
  {\bibinfo {volume} {108}},\ \bibinfo {pages} {3264} (\bibinfo {year}
  {1998})}\BibitemShut {NoStop}%
\bibitem [{\citenamefont {Appignanesi}\ \emph {et~al.}(2009)\citenamefont
  {Appignanesi}, \citenamefont {{Rodriguez Fris}},\ and\ \citenamefont
  {Sciortino}}]{Appignanesi2009}%
  \BibitemOpen
  \bibfield  {author} {\bibinfo {author} {\bibfnamefont {G.~A.}\ \bibnamefont
  {Appignanesi}}, \bibinfo {author} {\bibfnamefont {J.~A.}\ \bibnamefont
  {{Rodriguez Fris}}}, \ and\ \bibinfo {author} {\bibfnamefont
  {F.}~\bibnamefont {Sciortino}},\ }\href@noop {} {\bibfield  {journal}
  {\bibinfo  {journal}
  {\href{http://dx.doi.org/10.1140/epje/i2009-10478-6}{Eur. Phys. J. E.}}\
  }\textbf {\bibinfo {volume} {29}},\ \bibinfo {pages} {305} (\bibinfo {year}
  {2009})}\BibitemShut {NoStop}%
\bibitem [{\citenamefont {Accordino}\ \emph {et~al.}(2011)\citenamefont
  {Accordino}, \citenamefont {{Rodriguez Fris}}, \citenamefont {Sciortino},\
  and\ \citenamefont {Appignanesi}}]{Accordino2011}%
  \BibitemOpen
  \bibfield  {author} {\bibinfo {author} {\bibfnamefont {S.~R.}\ \bibnamefont
  {Accordino}}, \bibinfo {author} {\bibfnamefont {J.~A.}\ \bibnamefont
  {{Rodriguez Fris}}}, \bibinfo {author} {\bibfnamefont {F.}~\bibnamefont
  {Sciortino}}, \ and\ \bibinfo {author} {\bibfnamefont {G.~A.}\ \bibnamefont
  {Appignanesi}},\ }\href@noop {} {\bibfield  {journal} {\bibinfo  {journal}
  {\href{http://dx.doi.org/10.1140/epje/i2011-11048-1}{Eur. Phys. J. E.}}\
  }\textbf {\bibinfo {volume} {34}},\ \bibinfo {pages} {1} (\bibinfo {year}
  {2011})}\BibitemShut {NoStop}%
\bibitem [{\citenamefont {Wikfeldt}\ \emph {et~al.}(2011)\citenamefont
  {Wikfeldt}, \citenamefont {Nilsson},\ and\ \citenamefont
  {Pettersson}}]{Wikfeldt2011}%
  \BibitemOpen
  \bibfield  {author} {\bibinfo {author} {\bibfnamefont {K.~T.}\ \bibnamefont
  {Wikfeldt}}, \bibinfo {author} {\bibfnamefont {A.}~\bibnamefont {Nilsson}}, \
  and\ \bibinfo {author} {\bibfnamefont {L.~G.~M.}\ \bibnamefont
  {Pettersson}},\ }\href@noop {} {\bibfield  {journal} {\bibinfo  {journal}
  {\href{http://dx.doi.org/10.1039/c1cp22076d}{Phys. Chem. Chem. Phys.}}\
  }\textbf {\bibinfo {volume} {13}},\ \bibinfo {pages} {19918} (\bibinfo {year}
  {2011})}\BibitemShut {NoStop}%
\bibitem [{\citenamefont {Clark}\ \emph {et~al.}(2010)\citenamefont {Clark},
  \citenamefont {Hura}, \citenamefont {Teixeira}, \citenamefont {Soper},\ and\
  \citenamefont {Head-Gordon}}]{Clark2010}%
  \BibitemOpen
  \bibfield  {author} {\bibinfo {author} {\bibfnamefont {G.~N.~I.}\
  \bibnamefont {Clark}}, \bibinfo {author} {\bibfnamefont {G.~L.}\ \bibnamefont
  {Hura}}, \bibinfo {author} {\bibfnamefont {J.}~\bibnamefont {Teixeira}},
  \bibinfo {author} {\bibfnamefont {A.~K.}\ \bibnamefont {Soper}}, \ and\
  \bibinfo {author} {\bibfnamefont {T.}~\bibnamefont {Head-Gordon}},\
  }\href@noop {} {\bibfield  {journal} {\bibinfo  {journal}
  {\href{http://dx.doi.org/10.1073/pnas.1006599107}{Proc. Natl. Acad. Sci.
  USA}}\ }\textbf {\bibinfo {volume} {107}},\ \bibinfo {pages} {14003}
  (\bibinfo {year} {2010})}\BibitemShut {NoStop}%
\bibitem [{\citenamefont {Sedlmeier}\ \emph {et~al.}(2011)\citenamefont
  {Sedlmeier}, \citenamefont {Horinek},\ and\ \citenamefont
  {Netz}}]{Sedlmeier2011}%
  \BibitemOpen
  \bibfield  {author} {\bibinfo {author} {\bibfnamefont {F.}~\bibnamefont
  {Sedlmeier}}, \bibinfo {author} {\bibfnamefont {D.}~\bibnamefont {Horinek}},
  \ and\ \bibinfo {author} {\bibfnamefont {R.~R.}\ \bibnamefont {Netz}},\
  }\href@noop {} {\bibfield  {journal} {\bibinfo  {journal}
  {\href{http://dx.doi.org/10.1021/ja1064137}{J. Am. Chem. Soc.}}\ }\textbf
  {\bibinfo {volume} {133}},\ \bibinfo {pages} {1391} (\bibinfo {year}
  {2011})}\BibitemShut {NoStop}%
\bibitem [{\citenamefont {English}\ and\ \citenamefont
  {Tse}(2011)}]{English2011}%
  \BibitemOpen
  \bibfield  {author} {\bibinfo {author} {\bibfnamefont {N.~J.}\ \bibnamefont
  {English}}\ and\ \bibinfo {author} {\bibfnamefont {J.~S.}\ \bibnamefont
  {Tse}},\ }\href@noop {} {\bibfield  {journal} {\bibinfo  {journal}
  {\href{http://dx.doi.org/10.1103/PhysRevLett.106.037801}{Phys. Rev. Lett.}}\
  }\textbf {\bibinfo {volume} {106}},\ \bibinfo {pages} {037801} (\bibinfo
  {year} {2011})}\BibitemShut {NoStop}%
\bibitem [{\citenamefont {Asthagiri}\ \emph {et~al.}(2003)\citenamefont
  {Asthagiri}, \citenamefont {Pratt},\ and\ \citenamefont
  {Kress}}]{asthagiri_pre_2003}%
  \BibitemOpen
  \bibfield  {author} {\bibinfo {author} {\bibfnamefont {D.}~\bibnamefont
  {Asthagiri}}, \bibinfo {author} {\bibfnamefont {L.~R.}\ \bibnamefont
  {Pratt}}, \ and\ \bibinfo {author} {\bibfnamefont {J.~D.}\ \bibnamefont
  {Kress}},\ }\href@noop {} {\bibfield  {journal} {\bibinfo  {journal}
  {\href{http://dx.doi.org/10.1103/PhysRevE.68.041505}{Phys. Rev. E}}\ }\textbf
  {\bibinfo {volume} {68}},\ \bibinfo {pages} {041505} (\bibinfo {year}
  {2003})}\BibitemShut {NoStop}%
\bibitem [{\citenamefont {Grossman}\ \emph {et~al.}(2004)\citenamefont
  {Grossman}, \citenamefont {Schwegler}, \citenamefont {Draeger}, \citenamefont
  {Gygi},\ and\ \citenamefont {Galli}}]{grossman_jcp_2004}%
  \BibitemOpen
  \bibfield  {author} {\bibinfo {author} {\bibfnamefont {J.~C.}\ \bibnamefont
  {Grossman}}, \bibinfo {author} {\bibfnamefont {E.}~\bibnamefont {Schwegler}},
  \bibinfo {author} {\bibfnamefont {E.~W.}\ \bibnamefont {Draeger}}, \bibinfo
  {author} {\bibfnamefont {F.}~\bibnamefont {Gygi}}, \ and\ \bibinfo {author}
  {\bibfnamefont {G.}~\bibnamefont {Galli}},\ }\href@noop {} {\bibfield
  {journal} {\bibinfo  {journal} {\href{http://dx.doi.org/10.1063/1.1630560}{J.
  Chem. Phys.}}\ }\textbf {\bibinfo {volume} {120}},\ \bibinfo {pages} {300}
  (\bibinfo {year} {2004})}\BibitemShut {NoStop}%
\bibitem [{\citenamefont {Schwegler}\ \emph {et~al.}(2004)\citenamefont
  {Schwegler}, \citenamefont {Grossman}, \citenamefont {Gygi},\ and\
  \citenamefont {Galli}}]{schwegler_jcp_2004}%
  \BibitemOpen
  \bibfield  {author} {\bibinfo {author} {\bibfnamefont {E.}~\bibnamefont
  {Schwegler}}, \bibinfo {author} {\bibfnamefont {J.~C.}\ \bibnamefont
  {Grossman}}, \bibinfo {author} {\bibfnamefont {F.}~\bibnamefont {Gygi}}, \
  and\ \bibinfo {author} {\bibfnamefont {G.}~\bibnamefont {Galli}},\
  }\href@noop {} {\bibfield  {journal} {\bibinfo  {journal}
  {\href{http://dx.doi.org/10.1063/1.1782074}{J. Chem. Phys.}}\ }\textbf
  {\bibinfo {volume} {121}},\ \bibinfo {pages} {5400} (\bibinfo {year}
  {2004})}\BibitemShut {NoStop}%
\bibitem [{\citenamefont {Fern\'{a}ndez-Serra}\ and\ \citenamefont
  {Artacho}(2004)}]{fernandez_jcp_2004}%
  \BibitemOpen
  \bibfield  {author} {\bibinfo {author} {\bibfnamefont {M.~V.}\ \bibnamefont
  {Fern\'{a}ndez-Serra}}\ and\ \bibinfo {author} {\bibfnamefont
  {E.}~\bibnamefont {Artacho}},\ }\href@noop {} {\bibfield  {journal} {\bibinfo
   {journal} {\href{http://dx.doi.org/10.1063/1.1813431}{J. Chem. Phys.}}\
  }\textbf {\bibinfo {volume} {121}},\ \bibinfo {pages} {11136} (\bibinfo
  {year} {2004})}\BibitemShut {NoStop}%
\bibitem [{\citenamefont {Kuo}\ \emph {et~al.}(2004)\citenamefont {Kuo},
  \citenamefont {Mundy}, \citenamefont {McGrath}, \citenamefont {Siepmann},
  \citenamefont {VandeVondele}, \citenamefont {Sprik}, \citenamefont {Hutter},
  \citenamefont {Chen}, \citenamefont {Klein}, \citenamefont {Mohamed},
  \citenamefont {Krack},\ and\ \citenamefont {Parrinello}}]{kuo_jpcb_2004}%
  \BibitemOpen
  \bibfield  {author} {\bibinfo {author} {\bibfnamefont {I.-F.~W.}\
  \bibnamefont {Kuo}}, \bibinfo {author} {\bibfnamefont {C.~J.}\ \bibnamefont
  {Mundy}}, \bibinfo {author} {\bibfnamefont {M.~J.}\ \bibnamefont {McGrath}},
  \bibinfo {author} {\bibfnamefont {J.~I.}\ \bibnamefont {Siepmann}}, \bibinfo
  {author} {\bibfnamefont {J.}~\bibnamefont {VandeVondele}}, \bibinfo {author}
  {\bibfnamefont {M.}~\bibnamefont {Sprik}}, \bibinfo {author} {\bibfnamefont
  {J.}~\bibnamefont {Hutter}}, \bibinfo {author} {\bibfnamefont
  {B.}~\bibnamefont {Chen}}, \bibinfo {author} {\bibfnamefont {M.~L.}\
  \bibnamefont {Klein}}, \bibinfo {author} {\bibfnamefont {F.}~\bibnamefont
  {Mohamed}}, \bibinfo {author} {\bibfnamefont {M.}~\bibnamefont {Krack}}, \
  and\ \bibinfo {author} {\bibfnamefont {M.}~\bibnamefont {Parrinello}},\
  }\href@noop {} {\bibfield  {journal} {\bibinfo  {journal}
  {\href{http://dx.doi.org/10.1021/jp047788i}{J. Phys. Chem. B}}\ }\textbf
  {\bibinfo {volume} {108}},\ \bibinfo {pages} {12990} (\bibinfo {year}
  {2004})}\BibitemShut {NoStop}%
\bibitem [{\citenamefont {McGrath}\ \emph {et~al.}(2005)\citenamefont
  {McGrath}, \citenamefont {Siepmann}, \citenamefont {Kuo}, \citenamefont
  {Mundy}, \citenamefont {VandeVondele}, \citenamefont {Hutter}, \citenamefont
  {Mohamed},\ and\ \citenamefont {Krack}}]{mcgrath_cpc_2005}%
  \BibitemOpen
  \bibfield  {author} {\bibinfo {author} {\bibfnamefont {M.~J.}\ \bibnamefont
  {McGrath}}, \bibinfo {author} {\bibfnamefont {J.~I.}\ \bibnamefont
  {Siepmann}}, \bibinfo {author} {\bibfnamefont {I.-F.~W.}\ \bibnamefont
  {Kuo}}, \bibinfo {author} {\bibfnamefont {C.~J.}\ \bibnamefont {Mundy}},
  \bibinfo {author} {\bibfnamefont {J.}~\bibnamefont {VandeVondele}}, \bibinfo
  {author} {\bibfnamefont {J.}~\bibnamefont {Hutter}}, \bibinfo {author}
  {\bibfnamefont {F.}~\bibnamefont {Mohamed}}, \ and\ \bibinfo {author}
  {\bibfnamefont {M.}~\bibnamefont {Krack}},\ }\href@noop {} {\bibfield
  {journal} {\bibinfo  {journal}
  {\href{http://dx.doi.org/10.1002/cphc.200400580}{ChemPhysChem}}\ }\textbf
  {\bibinfo {volume} {6}},\ \bibinfo {pages} {1894} (\bibinfo {year}
  {2005})}\BibitemShut {NoStop}%
\bibitem [{\citenamefont {VandeVondele}\ \emph {et~al.}(2005)\citenamefont
  {VandeVondele}, \citenamefont {Mohamed}, \citenamefont {Krack}, \citenamefont
  {Hutter}, \citenamefont {Sprik},\ and\ \citenamefont
  {Parrinello}}]{vandevondele_jcp_2005}%
  \BibitemOpen
  \bibfield  {author} {\bibinfo {author} {\bibfnamefont {J.}~\bibnamefont
  {VandeVondele}}, \bibinfo {author} {\bibfnamefont {F.}~\bibnamefont
  {Mohamed}}, \bibinfo {author} {\bibfnamefont {M.}~\bibnamefont {Krack}},
  \bibinfo {author} {\bibfnamefont {J.}~\bibnamefont {Hutter}}, \bibinfo
  {author} {\bibfnamefont {M.}~\bibnamefont {Sprik}}, \ and\ \bibinfo {author}
  {\bibfnamefont {M.}~\bibnamefont {Parrinello}},\ }\href@noop {} {\bibfield
  {journal} {\bibinfo  {journal} {\href{http://dx.doi.org/10.1063/1.1828433}{J.
  Chem. Phys.}}\ }\textbf {\bibinfo {volume} {122}},\ \bibinfo {pages} {014515}
  (\bibinfo {year} {2005})}\BibitemShut {NoStop}%
\bibitem [{\citenamefont {Sit}\ and\ \citenamefont
  {Marzari}(2005)}]{sit_jcp_2005}%
  \BibitemOpen
  \bibfield  {author} {\bibinfo {author} {\bibfnamefont {P.~H.-L.}\
  \bibnamefont {Sit}}\ and\ \bibinfo {author} {\bibfnamefont {N.}~\bibnamefont
  {Marzari}},\ }\href@noop {} {\bibfield  {journal} {\bibinfo  {journal}
  {\href{http://dx.doi.org/10.1063/1.1908913}{J. Chem. Phys.}}\ }\textbf
  {\bibinfo {volume} {122}},\ \bibinfo {pages} {204510} (\bibinfo {year}
  {2005})}\BibitemShut {NoStop}%
\bibitem [{\citenamefont {Fern\'{a}ndez-Serra}\ \emph
  {et~al.}(2005)\citenamefont {Fern\'{a}ndez-Serra}, \citenamefont {Ferlat},\
  and\ \citenamefont {Artacho}}]{fernandez_ms_2005}%
  \BibitemOpen
  \bibfield  {author} {\bibinfo {author} {\bibfnamefont {M.~V.}\ \bibnamefont
  {Fern\'{a}ndez-Serra}}, \bibinfo {author} {\bibfnamefont {G.}~\bibnamefont
  {Ferlat}}, \ and\ \bibinfo {author} {\bibfnamefont {E.}~\bibnamefont
  {Artacho}},\ }\href@noop {} {\bibfield  {journal} {\bibinfo  {journal}
  {\href{http://dx.doi.org/10.1080/08927020500066726}{Molecular Simualtion}}\
  }\textbf {\bibinfo {volume} {31}},\ \bibinfo {pages} {361} (\bibinfo {year}
  {2005})}\BibitemShut {NoStop}%
\bibitem [{\citenamefont {McGrath}\ \emph {et~al.}(2006)\citenamefont
  {McGrath}, \citenamefont {Siepmann}, \citenamefont {Kuo},\ and\ \citenamefont
  {Mundy}}]{mcgrath_mp_2006}%
  \BibitemOpen
  \bibfield  {author} {\bibinfo {author} {\bibfnamefont {M.~J.}\ \bibnamefont
  {McGrath}}, \bibinfo {author} {\bibfnamefont {J.~I.}\ \bibnamefont
  {Siepmann}}, \bibinfo {author} {\bibfnamefont {I.-F.~W.}\ \bibnamefont
  {Kuo}}, \ and\ \bibinfo {author} {\bibfnamefont {C.~J.}\ \bibnamefont
  {Mundy}},\ }\href@noop {} {\bibfield  {journal} {\bibinfo  {journal}
  {\href{http://dx.doi.org/10.1080/00268970601014781}{Mol. Phys.}}\ }\textbf
  {\bibinfo {volume} {104}},\ \bibinfo {pages} {3619} (\bibinfo {year}
  {2006})}\BibitemShut {NoStop}%
\bibitem [{\citenamefont {Lee}\ and\ \citenamefont
  {Tuckerman}(2006)}]{lee_jcp_2006}%
  \BibitemOpen
  \bibfield  {author} {\bibinfo {author} {\bibfnamefont {H.-S.}\ \bibnamefont
  {Lee}}\ and\ \bibinfo {author} {\bibfnamefont {M.~E.}\ \bibnamefont
  {Tuckerman}},\ }\href@noop {} {\bibfield  {journal} {\bibinfo  {journal}
  {\href{http://dx.doi.org/10.1063/1.2354158}{J. Chem. Phys.}}\ }\textbf
  {\bibinfo {volume} {125}},\ \bibinfo {pages} {154507} (\bibinfo {year}
  {2006})}\BibitemShut {NoStop}%
\bibitem [{\citenamefont {Todorova}\ \emph {et~al.}(2006)\citenamefont
  {Todorova}, \citenamefont {Seitsonen}, \citenamefont {Hutter}, \citenamefont
  {Kuo},\ and\ \citenamefont {Mundy}}]{todorova_jpcb_2006}%
  \BibitemOpen
  \bibfield  {author} {\bibinfo {author} {\bibfnamefont {T.}~\bibnamefont
  {Todorova}}, \bibinfo {author} {\bibfnamefont {A.~P.}\ \bibnamefont
  {Seitsonen}}, \bibinfo {author} {\bibfnamefont {J.}~\bibnamefont {Hutter}},
  \bibinfo {author} {\bibfnamefont {I.-F.~W.}\ \bibnamefont {Kuo}}, \ and\
  \bibinfo {author} {\bibfnamefont {C.~J.}\ \bibnamefont {Mundy}},\ }\href@noop
  {} {\bibfield  {journal} {\bibinfo  {journal}
  {\href{http://dx.doi.org/10.1021/jp055127v}{J. Phys. Chem. B}}\ }\textbf
  {\bibinfo {volume} {110}},\ \bibinfo {pages} {3685} (\bibinfo {year}
  {2006})}\BibitemShut {NoStop}%
\bibitem [{\citenamefont {Lee}\ and\ \citenamefont
  {Tuckerman}(2007)}]{lee_jcp_2007}%
  \BibitemOpen
  \bibfield  {author} {\bibinfo {author} {\bibfnamefont {H.-S.}\ \bibnamefont
  {Lee}}\ and\ \bibinfo {author} {\bibfnamefont {M.~E.}\ \bibnamefont
  {Tuckerman}},\ }\href@noop {} {\bibfield  {journal} {\bibinfo  {journal}
  {\href{http://dx.doi.org/10.1063/1.2718521}{J. Chem. Phys.}}\ }\textbf
  {\bibinfo {volume} {126}},\ \bibinfo {pages} {164501} (\bibinfo {year}
  {2007})}\BibitemShut {NoStop}%
\bibitem [{\citenamefont {Guidon}\ \emph {et~al.}(2008)\citenamefont {Guidon},
  \citenamefont {Schiffmann}, \citenamefont {Hutter},\ and\ \citenamefont
  {VandeVondele}}]{guidon_jcp_2008}%
  \BibitemOpen
  \bibfield  {author} {\bibinfo {author} {\bibfnamefont {M.}~\bibnamefont
  {Guidon}}, \bibinfo {author} {\bibfnamefont {F.}~\bibnamefont {Schiffmann}},
  \bibinfo {author} {\bibfnamefont {J.}~\bibnamefont {Hutter}}, \ and\ \bibinfo
  {author} {\bibfnamefont {J.}~\bibnamefont {VandeVondele}},\ }\href@noop {}
  {\bibfield  {journal} {\bibinfo  {journal}
  {\href{http://dx.doi.org/10.1063/1.2931945}{J. Chem. Phys.}}\ }\textbf
  {\bibinfo {volume} {128}},\ \bibinfo {pages} {214104} (\bibinfo {year}
  {2008})}\BibitemShut {NoStop}%
\bibitem [{\citenamefont {K\"{u}hne}\ \emph {et~al.}(2009)\citenamefont
  {K\"{u}hne}, \citenamefont {Krack},\ and\ \citenamefont
  {Parrinello}}]{kuhne_jctc_2009}%
  \BibitemOpen
  \bibfield  {author} {\bibinfo {author} {\bibfnamefont {T.~D.}\ \bibnamefont
  {K\"{u}hne}}, \bibinfo {author} {\bibfnamefont {M.}~\bibnamefont {Krack}}, \
  and\ \bibinfo {author} {\bibfnamefont {M.}~\bibnamefont {Parrinello}},\
  }\href@noop {} {\bibfield  {journal} {\bibinfo  {journal}
  {\href{http://dx.doi.org/10.1021/ct800417q}{J. Chem. Theory Comput.}}\
  }\textbf {\bibinfo {volume} {5}},\ \bibinfo {pages} {235} (\bibinfo {year}
  {2009})}\BibitemShut {NoStop}%
\bibitem [{\citenamefont {Mattson}\ and\ \citenamefont
  {Mattson}(2009)}]{mattson_jctc_2009}%
  \BibitemOpen
  \bibfield  {author} {\bibinfo {author} {\bibfnamefont {A.~E.}\ \bibnamefont
  {Mattson}}\ and\ \bibinfo {author} {\bibfnamefont {T.~R.}\ \bibnamefont
  {Mattson}},\ }\href@noop {} {\bibfield  {journal} {\bibinfo  {journal}
  {\href{http://dx.doi.org/10.1021/ct8004968}{J. Chem. Theory Comput.}}\
  }\textbf {\bibinfo {volume} {5}},\ \bibinfo {pages} {887} (\bibinfo {year}
  {2009})}\BibitemShut {NoStop}%
\bibitem [{\citenamefont {Yoo}\ \emph {et~al.}(2009)\citenamefont {Yoo},
  \citenamefont {Zeng},\ and\ \citenamefont {Xantheas}}]{yoo_jcp_2009}%
  \BibitemOpen
  \bibfield  {author} {\bibinfo {author} {\bibfnamefont {S.}~\bibnamefont
  {Yoo}}, \bibinfo {author} {\bibfnamefont {X.~C.}\ \bibnamefont {Zeng}}, \
  and\ \bibinfo {author} {\bibfnamefont {S.~S.}\ \bibnamefont {Xantheas}},\
  }\href@noop {} {\bibfield  {journal} {\bibinfo  {journal}
  {\href{http://dx.doi.org/10.1063/1.3153871}{J. Chem. Phys.}}\ }\textbf
  {\bibinfo {volume} {130}},\ \bibinfo {pages} {221102} (\bibinfo {year}
  {2009})}\BibitemShut {NoStop}%
\bibitem [{\citenamefont {Zhang}\ \emph
  {et~al.}(2011{\natexlab{a}})\citenamefont {Zhang}, \citenamefont {Donadio},
  \citenamefont {Gygi},\ and\ \citenamefont {Galli}}]{zhang_jctc_2011-2}%
  \BibitemOpen
  \bibfield  {author} {\bibinfo {author} {\bibfnamefont {C.}~\bibnamefont
  {Zhang}}, \bibinfo {author} {\bibfnamefont {D.}~\bibnamefont {Donadio}},
  \bibinfo {author} {\bibfnamefont {F.}~\bibnamefont {Gygi}}, \ and\ \bibinfo
  {author} {\bibfnamefont {G.}~\bibnamefont {Galli}},\ }\href@noop {}
  {\bibfield  {journal} {\bibinfo  {journal}
  {\href{http://dx.doi.org/10.1021/ct2000952}{J. Chem. Theory Comput.}}\
  }\textbf {\bibinfo {volume} {7}},\ \bibinfo {pages} {1443} (\bibinfo {year}
  {2011}{\natexlab{a}})}\BibitemShut {NoStop}%
\bibitem [{\citenamefont {Bart\'{o}k}\ \emph {et~al.}(2013)\citenamefont
  {Bart\'{o}k}, \citenamefont {Gillan}, \citenamefont {Manby},\ and\
  \citenamefont {Cs\'{a}nyi}}]{Bartok2013}%
  \BibitemOpen
  \bibfield  {author} {\bibinfo {author} {\bibfnamefont {A.~P.}\ \bibnamefont
  {Bart\'{o}k}}, \bibinfo {author} {\bibfnamefont {M.~J.}\ \bibnamefont
  {Gillan}}, \bibinfo {author} {\bibfnamefont {F.~R.}\ \bibnamefont {Manby}}, \
  and\ \bibinfo {author} {\bibfnamefont {G.}~\bibnamefont {Cs\'{a}nyi}},\
  }\href@noop {} {\bibfield  {journal} {\bibinfo  {journal}
  {\href{http://dx.doi.org/10.1103/PhysRevB.88.054104}{Phys. Rev. B}}\ }\textbf
  {\bibinfo {volume} {88}},\ \bibinfo {pages} {054104} (\bibinfo {year}
  {2013})}\BibitemShut {NoStop}%
\bibitem [{\citenamefont {Alf\`{e}}\ \emph {et~al.}(2013)\citenamefont
  {Alf\`{e}}, \citenamefont {Bart\'{o}k}, \citenamefont {Cs\'{a}nyi},\ and\
  \citenamefont {Gillan}}]{Alfe2013}%
  \BibitemOpen
  \bibfield  {author} {\bibinfo {author} {\bibfnamefont {D.}~\bibnamefont
  {Alf\`{e}}}, \bibinfo {author} {\bibfnamefont {A.~P.}\ \bibnamefont
  {Bart\'{o}k}}, \bibinfo {author} {\bibfnamefont {G.}~\bibnamefont
  {Cs\'{a}nyi}}, \ and\ \bibinfo {author} {\bibfnamefont {M.~J.}\ \bibnamefont
  {Gillan}},\ }\href@noop {} {\bibfield  {journal} {\bibinfo  {journal}
  {\href{http://dx.doi.org/10.1063/1.4810882}{J. Chem. Phys.}}\ }\textbf
  {\bibinfo {volume} {138}},\ \bibinfo {pages} {221102} (\bibinfo {year}
  {2013})}\BibitemShut {NoStop}%
\bibitem [{\citenamefont {Lin}\ \emph {et~al.}(2009)\citenamefont {Lin},
  \citenamefont {Seitsonen}, \citenamefont {Coutinho-Neto}, \citenamefont
  {Tavernelli},\ and\ \citenamefont {Rothlisberger}}]{lin_jpcb_2009}%
  \BibitemOpen
  \bibfield  {author} {\bibinfo {author} {\bibfnamefont {I.-C.}\ \bibnamefont
  {Lin}}, \bibinfo {author} {\bibfnamefont {A.~P.}\ \bibnamefont {Seitsonen}},
  \bibinfo {author} {\bibfnamefont {M.~D.}\ \bibnamefont {Coutinho-Neto}},
  \bibinfo {author} {\bibfnamefont {I.}~\bibnamefont {Tavernelli}}, \ and\
  \bibinfo {author} {\bibfnamefont {U.}~\bibnamefont {Rothlisberger}},\
  }\href@noop {} {\bibfield  {journal} {\bibinfo  {journal}
  {\href{http://dx.doi.org/10.1021/jp806376e}{J. Phys. Chem. B}}\ }\textbf
  {\bibinfo {volume} {113}},\ \bibinfo {pages} {1127} (\bibinfo {year}
  {2009})}\BibitemShut {NoStop}%
\bibitem [{\citenamefont {Jonchiere}\ \emph {et~al.}(2011)\citenamefont
  {Jonchiere}, \citenamefont {Seitsonen}, \citenamefont {Ferlat}, \citenamefont
  {Saitta},\ and\ \citenamefont {Vuilleumier}}]{jonchiere_jcp_2011}%
  \BibitemOpen
  \bibfield  {author} {\bibinfo {author} {\bibfnamefont {R.}~\bibnamefont
  {Jonchiere}}, \bibinfo {author} {\bibfnamefont {A.~P.}\ \bibnamefont
  {Seitsonen}}, \bibinfo {author} {\bibfnamefont {G.}~\bibnamefont {Ferlat}},
  \bibinfo {author} {\bibfnamefont {A.~M.}\ \bibnamefont {Saitta}}, \ and\
  \bibinfo {author} {\bibfnamefont {R.}~\bibnamefont {Vuilleumier}},\
  }\href@noop {} {\bibfield  {journal} {\bibinfo  {journal}
  {\href{http://dx.doi.org/10.1063/1.3651474}{J. Chem. Phys.}}\ }\textbf
  {\bibinfo {volume} {135}},\ \bibinfo {pages} {154503} (\bibinfo {year}
  {2011})}\BibitemShut {NoStop}%
\bibitem [{\citenamefont {Wang}\ \emph {et~al.}(2011)\citenamefont {Wang},
  \citenamefont {Rom\'{a}n-P\'{e}rez}, \citenamefont {Soler}, \citenamefont
  {Artacho},\ and\ \citenamefont {Fern\'{a}ndez-Serra}}]{wang_jcp_2011}%
  \BibitemOpen
  \bibfield  {author} {\bibinfo {author} {\bibfnamefont {J.}~\bibnamefont
  {Wang}}, \bibinfo {author} {\bibfnamefont {G.}~\bibnamefont
  {Rom\'{a}n-P\'{e}rez}}, \bibinfo {author} {\bibfnamefont {J.~M.}\
  \bibnamefont {Soler}}, \bibinfo {author} {\bibfnamefont {E.}~\bibnamefont
  {Artacho}}, \ and\ \bibinfo {author} {\bibfnamefont {M.-V.}\ \bibnamefont
  {Fern\'{a}ndez-Serra}},\ }\href@noop {} {\bibfield  {journal} {\bibinfo
  {journal} {\href{http://dx.doi.org/10.1063/1.3521268}{J. Chem. Phys.}}\
  }\textbf {\bibinfo {volume} {134}},\ \bibinfo {pages} {024516} (\bibinfo
  {year} {2011})}\BibitemShut {NoStop}%
\bibitem [{\citenamefont {Zhang}\ \emph
  {et~al.}(2011{\natexlab{b}})\citenamefont {Zhang}, \citenamefont {Wu},
  \citenamefont {Galli},\ and\ \citenamefont {Gygi}}]{zhang_jctc_2011}%
  \BibitemOpen
  \bibfield  {author} {\bibinfo {author} {\bibfnamefont {C.}~\bibnamefont
  {Zhang}}, \bibinfo {author} {\bibfnamefont {J.}~\bibnamefont {Wu}}, \bibinfo
  {author} {\bibfnamefont {G.}~\bibnamefont {Galli}}, \ and\ \bibinfo {author}
  {\bibfnamefont {F.}~\bibnamefont {Gygi}},\ }\href@noop {} {\bibfield
  {journal} {\bibinfo  {journal} {\href{http://dx.doi.org/10.1021/ct200329e}{J.
  Chem. Theory Comput.}}\ }\textbf {\bibinfo {volume} {7}},\ \bibinfo {pages}
  {3054} (\bibinfo {year} {2011}{\natexlab{b}})}\BibitemShut {NoStop}%
\bibitem [{\citenamefont {M{\o}gelh{\o}j}\ \emph {et~al.}(2011)\citenamefont
  {M{\o}gelh{\o}j}, \citenamefont {Kelkkanen}, \citenamefont {Wikfeldt},
  \citenamefont {Schi{\o}tz}, \citenamefont {Mortensen}, \citenamefont
  {Pettersson}, \citenamefont {Lundqvist}, \citenamefont {Jacobsen},
  \citenamefont {Nilsson},\ and\ \citenamefont
  {N{\o}rskov}}]{mogelhoj_jpcb_2011}%
  \BibitemOpen
  \bibfield  {author} {\bibinfo {author} {\bibfnamefont {A.}~\bibnamefont
  {M{\o}gelh{\o}j}}, \bibinfo {author} {\bibfnamefont {A.~K.}\ \bibnamefont
  {Kelkkanen}}, \bibinfo {author} {\bibfnamefont {K.~T.}\ \bibnamefont
  {Wikfeldt}}, \bibinfo {author} {\bibfnamefont {J.}~\bibnamefont
  {Schi{\o}tz}}, \bibinfo {author} {\bibfnamefont {J.~J.}\ \bibnamefont
  {Mortensen}}, \bibinfo {author} {\bibfnamefont {L.~G.~M.}\ \bibnamefont
  {Pettersson}}, \bibinfo {author} {\bibfnamefont {B.~I.}\ \bibnamefont
  {Lundqvist}}, \bibinfo {author} {\bibfnamefont {K.~W.}\ \bibnamefont
  {Jacobsen}}, \bibinfo {author} {\bibfnamefont {A.}~\bibnamefont {Nilsson}}, \
  and\ \bibinfo {author} {\bibfnamefont {J.~K.}\ \bibnamefont {N{\o}rskov}},\
  }\href@noop {} {\bibfield  {journal} {\bibinfo  {journal}
  {\href{http://dx.doi.org/10.1021/jp2040345}{J. Phys. Chem. B}}\ }\textbf
  {\bibinfo {volume} {115}},\ \bibinfo {pages} {14149} (\bibinfo {year}
  {2011})}\BibitemShut {NoStop}%
\bibitem [{\citenamefont {Lin}\ \emph {et~al.}(2012)\citenamefont {Lin},
  \citenamefont {Seitsonen}, \citenamefont {Tavernelli},\ and\ \citenamefont
  {Rothlisberger}}]{lin_jctc_2012}%
  \BibitemOpen
  \bibfield  {author} {\bibinfo {author} {\bibfnamefont {I.-C.}\ \bibnamefont
  {Lin}}, \bibinfo {author} {\bibfnamefont {A.~P.}\ \bibnamefont {Seitsonen}},
  \bibinfo {author} {\bibfnamefont {I.}~\bibnamefont {Tavernelli}}, \ and\
  \bibinfo {author} {\bibfnamefont {U.}~\bibnamefont {Rothlisberger}},\
  }\href@noop {} {\bibfield  {journal} {\bibinfo  {journal}
  {\href{http://dx.doi.org/10.1021/ct3001848}{J. Chem. Theory Comput.}}\
  }\textbf {\bibinfo {volume} {8}},\ \bibinfo {pages} {3902} (\bibinfo {year}
  {2012})}\BibitemShut {NoStop}%
\bibitem [{\citenamefont {Yoo}\ and\ \citenamefont
  {Xantheas}(2011)}]{yoo_jcp_2011}%
  \BibitemOpen
  \bibfield  {author} {\bibinfo {author} {\bibfnamefont {S.}~\bibnamefont
  {Yoo}}\ and\ \bibinfo {author} {\bibfnamefont {S.~S.}\ \bibnamefont
  {Xantheas}},\ }\href@noop {} {\bibfield  {journal} {\bibinfo  {journal}
  {\href{http://dx.doi.org/10.1063/1.3573375}{J. Chem. Phys.}}\ }\textbf
  {\bibinfo {volume} {134}},\ \bibinfo {pages} {121105} (\bibinfo {year}
  {2011})}\BibitemShut {NoStop}%
\bibitem [{\citenamefont {Schmidt}\ \emph {et~al.}(2009)\citenamefont
  {Schmidt}, \citenamefont {VandeVondele}, \citenamefont {Kuo}, \citenamefont
  {Sebastiani}, \citenamefont {Siepmann}, \citenamefont {Hutter},\ and\
  \citenamefont {Mundy}}]{schmidt_jpcb_2009}%
  \BibitemOpen
  \bibfield  {author} {\bibinfo {author} {\bibfnamefont {J.}~\bibnamefont
  {Schmidt}}, \bibinfo {author} {\bibfnamefont {J.}~\bibnamefont
  {VandeVondele}}, \bibinfo {author} {\bibfnamefont {I.-F.~W.}\ \bibnamefont
  {Kuo}}, \bibinfo {author} {\bibfnamefont {D.}~\bibnamefont {Sebastiani}},
  \bibinfo {author} {\bibfnamefont {J.~I.}\ \bibnamefont {Siepmann}}, \bibinfo
  {author} {\bibfnamefont {J.}~\bibnamefont {Hutter}}, \ and\ \bibinfo {author}
  {\bibfnamefont {C.~J.}\ \bibnamefont {Mundy}},\ }\href@noop {} {\bibfield
  {journal} {\bibinfo  {journal} {\href{http://dx.doi.org/10.1021/jp901990u}{J.
  Phys. Chem. B}}\ }\textbf {\bibinfo {volume} {113}},\ \bibinfo {pages}
  {11959} (\bibinfo {year} {2009})}\BibitemShut {NoStop}%
\bibitem [{\citenamefont {Ma}\ \emph {et~al.}(2012)\citenamefont {Ma},
  \citenamefont {Zhang},\ and\ \citenamefont {Tuckerman}}]{ma_jcp_2012}%
  \BibitemOpen
  \bibfield  {author} {\bibinfo {author} {\bibfnamefont {Z.}~\bibnamefont
  {Ma}}, \bibinfo {author} {\bibfnamefont {Y.}~\bibnamefont {Zhang}}, \ and\
  \bibinfo {author} {\bibfnamefont {M.~E.}\ \bibnamefont {Tuckerman}},\
  }\href@noop {} {\bibfield  {journal} {\bibinfo  {journal}
  {\href{http://dx.doi.org/10.1063/1.4736712}{J. Chem. Phys.}}\ }\textbf
  {\bibinfo {volume} {137}},\ \bibinfo {pages} {044506} (\bibinfo {year}
  {2012})}\BibitemShut {NoStop}%
\bibitem [{\citenamefont {Ben}\ \emph {et~al.}(2013)\citenamefont {Ben},
  \citenamefont {Sch\"{o}ner}, \citenamefont {Hutter},\ and\ \citenamefont
  {VandeVondele}}]{ben_jpcl_2013}%
  \BibitemOpen
  \bibfield  {author} {\bibinfo {author} {\bibfnamefont {M.~D.}\ \bibnamefont
  {Ben}}, \bibinfo {author} {\bibfnamefont {M.}~\bibnamefont {Sch\"{o}ner}},
  \bibinfo {author} {\bibfnamefont {J.}~\bibnamefont {Hutter}}, \ and\ \bibinfo
  {author} {\bibfnamefont {J.}~\bibnamefont {VandeVondele}},\ }\href@noop {}
  {\bibfield  {journal} {\bibinfo  {journal}
  {\href{http://dx.doi.org/10.1021/jz401931f}{J. Phys. Chem. Lett.}}\ }\textbf
  {\bibinfo {volume} {4}},\ \bibinfo {pages} {3753} (\bibinfo {year}
  {2013})}\BibitemShut {NoStop}%
\bibitem [{\citenamefont {Corsetti}\ \emph {et~al.}(2013)\citenamefont
  {Corsetti}, \citenamefont {Artacho}, \citenamefont {Soler}, \citenamefont
  {Alexandre},\ and\ \citenamefont {Fern\'{a}ndez-Serra}}]{Corsetti2013}%
  \BibitemOpen
  \bibfield  {author} {\bibinfo {author} {\bibfnamefont {F.}~\bibnamefont
  {Corsetti}}, \bibinfo {author} {\bibfnamefont {E.}~\bibnamefont {Artacho}},
  \bibinfo {author} {\bibfnamefont {J.~M.}\ \bibnamefont {Soler}}, \bibinfo
  {author} {\bibfnamefont {S.~S.}\ \bibnamefont {Alexandre}}, \ and\ \bibinfo
  {author} {\bibfnamefont {M.-V.}\ \bibnamefont {Fern\'{a}ndez-Serra}},\
  }\href@noop {} {\bibfield  {journal} {\bibinfo  {journal}
  {\href{http://dx.doi.org/10.1063/1.4832141}{J. Chem. Phys.}}\ }\textbf
  {\bibinfo {volume} {139}},\ \bibinfo {pages} {194502} (\bibinfo {year}
  {2013})}\BibitemShut {NoStop}%
\bibitem [{\citenamefont {Bankura}\ \emph {et~al.}(2014)\citenamefont
  {Bankura}, \citenamefont {Karmakar}, \citenamefont {Carnevale}, \citenamefont
  {Chandra},\ and\ \citenamefont {Klein}}]{Bankura2014}%
  \BibitemOpen
  \bibfield  {author} {\bibinfo {author} {\bibfnamefont {A.}~\bibnamefont
  {Bankura}}, \bibinfo {author} {\bibfnamefont {A.}~\bibnamefont {Karmakar}},
  \bibinfo {author} {\bibfnamefont {V.}~\bibnamefont {Carnevale}}, \bibinfo
  {author} {\bibfnamefont {A.}~\bibnamefont {Chandra}}, \ and\ \bibinfo
  {author} {\bibfnamefont {M.~L.}\ \bibnamefont {Klein}},\ }\href@noop {}
  {\bibfield  {journal} {\bibinfo  {journal}
  {\href{http://dx.doi.org/10.1021/jp506120t}{J. Phys. Chem. C}}\ }\textbf
  {\bibinfo {volume} {118}},\ \bibinfo {pages} {29401} (\bibinfo {year}
  {2014})}\BibitemShut {NoStop}%
\bibitem [{\citenamefont {Forster-Tonigold}\ and\ \citenamefont
  {Gro\ss}(2014)}]{Forster-Tonigold2014}%
  \BibitemOpen
  \bibfield  {author} {\bibinfo {author} {\bibfnamefont {K.}~\bibnamefont
  {Forster-Tonigold}}\ and\ \bibinfo {author} {\bibfnamefont {A.}~\bibnamefont
  {Gro\ss}},\ }\href@noop {} {\bibfield  {journal} {\bibinfo  {journal}
  {\href{http://dx.doi.org/10.1063/1.4892400}{J. Chem. Phys.}}\ }\textbf
  {\bibinfo {volume} {141}},\ \bibinfo {pages} {064501} (\bibinfo {year}
  {2014})}\BibitemShut {NoStop}%
\bibitem [{\citenamefont {Miceli}\ \emph {et~al.}(2015)\citenamefont {Miceli},
  \citenamefont {Gironcoli},\ and\ \citenamefont {Pasquarello}}]{Miceli2015}%
  \BibitemOpen
  \bibfield  {author} {\bibinfo {author} {\bibfnamefont {G.}~\bibnamefont
  {Miceli}}, \bibinfo {author} {\bibfnamefont {S.~D.}\ \bibnamefont
  {Gironcoli}}, \ and\ \bibinfo {author} {\bibfnamefont {A.}~\bibnamefont
  {Pasquarello}},\ }\href@noop {} {\bibfield  {journal} {\bibinfo  {journal}
  {\href{http://dx.doi.org/10.1063/1.4905333}{J. Chem. Phys.}}\ }\textbf
  {\bibinfo {volume} {142}},\ \bibinfo {pages} {034501} (\bibinfo {year}
  {2015})}\BibitemShut {NoStop}%
\bibitem [{\citenamefont {Perdew}\ and\ \citenamefont
  {Zunger}(1981)}]{perdew_prb_1981}%
  \BibitemOpen
  \bibfield  {author} {\bibinfo {author} {\bibfnamefont {J.}~\bibnamefont
  {Perdew}}\ and\ \bibinfo {author} {\bibfnamefont {A.}~\bibnamefont
  {Zunger}},\ }\href@noop {} {\bibfield  {journal} {\bibinfo  {journal}
  {\href{http://dx.doi.org/10.1103/PhysRevB.23.5048}{Phys. Rev. B}}\ }\textbf
  {\bibinfo {volume} {23}},\ \bibinfo {pages} {5048} (\bibinfo {year}
  {1981})}\BibitemShut {NoStop}%
\bibitem [{\citenamefont {Perdew}\ \emph
  {et~al.}(1996{\natexlab{a}})\citenamefont {Perdew}, \citenamefont
  {Ernzerhof},\ and\ \citenamefont {Burke}}]{perdew_jcp_1996}%
  \BibitemOpen
  \bibfield  {author} {\bibinfo {author} {\bibfnamefont {J.~P.}\ \bibnamefont
  {Perdew}}, \bibinfo {author} {\bibfnamefont {M.}~\bibnamefont {Ernzerhof}}, \
  and\ \bibinfo {author} {\bibfnamefont {K.}~\bibnamefont {Burke}},\
  }\href@noop {} {\bibfield  {journal} {\bibinfo  {journal}
  {\href{http://dx.doi.org/10.1063/1.472933}{J. Chem. Phys.}}\ }\textbf
  {\bibinfo {volume} {105}},\ \bibinfo {pages} {9982} (\bibinfo {year}
  {1996}{\natexlab{a}})}\BibitemShut {NoStop}%
\bibitem [{\citenamefont {Adamo}\ and\ \citenamefont
  {Barone}(1999)}]{adamo_jcp_1999}%
  \BibitemOpen
  \bibfield  {author} {\bibinfo {author} {\bibfnamefont {C.}~\bibnamefont
  {Adamo}}\ and\ \bibinfo {author} {\bibfnamefont {V.}~\bibnamefont {Barone}},\
  }\href@noop {} {\bibfield  {journal} {\bibinfo  {journal}
  {\href{http://dx.doi.org/10.1063/1.478522}{J. Chem. Phys.}}\ }\textbf
  {\bibinfo {volume} {110}},\ \bibinfo {pages} {6158} (\bibinfo {year}
  {1999})}\BibitemShut {NoStop}%
\bibitem [{\citenamefont {Tkatchenko}\ and\ \citenamefont
  {Scheffler}(2009)}]{tkatchenko_prl_2009}%
  \BibitemOpen
  \bibfield  {author} {\bibinfo {author} {\bibfnamefont {A.}~\bibnamefont
  {Tkatchenko}}\ and\ \bibinfo {author} {\bibfnamefont {M.}~\bibnamefont
  {Scheffler}},\ }\href@noop {} {\bibfield  {journal} {\bibinfo  {journal}
  {\href{http://dx.doi.org/10.1103/PhysRevLett.102.073005}{Phys. Rev. Lett.}}\
  }\textbf {\bibinfo {volume} {102}},\ \bibinfo {pages} {073005} (\bibinfo
  {year} {2009})}\BibitemShut {NoStop}%
\bibitem [{\citenamefont {Skinner}\ \emph {et~al.}(2013)\citenamefont
  {Skinner}, \citenamefont {Huang}, \citenamefont {Schlesinger}, \citenamefont
  {Pettersson}, \citenamefont {Nilsson},\ and\ \citenamefont
  {Benmore}}]{Skinner2013}%
  \BibitemOpen
  \bibfield  {author} {\bibinfo {author} {\bibfnamefont {L.~B.}\ \bibnamefont
  {Skinner}}, \bibinfo {author} {\bibfnamefont {C.}~\bibnamefont {Huang}},
  \bibinfo {author} {\bibfnamefont {D.}~\bibnamefont {Schlesinger}}, \bibinfo
  {author} {\bibfnamefont {L.~G.~M.}\ \bibnamefont {Pettersson}}, \bibinfo
  {author} {\bibfnamefont {A.}~\bibnamefont {Nilsson}}, \ and\ \bibinfo
  {author} {\bibfnamefont {C.~J.}\ \bibnamefont {Benmore}},\ }\href@noop {}
  {\bibfield  {journal} {\bibinfo  {journal}
  {\href{http://dx.doi.org/10.1063/1.4790861}{J. Chem. Phys.}}\ }\textbf
  {\bibinfo {volume} {138}},\ \bibinfo {pages} {074506} (\bibinfo {year}
  {2013})}\BibitemShut {NoStop}%
\bibitem [{\citenamefont {Car}\ and\ \citenamefont
  {Parrinello}(1985)}]{car_prl_1985}%
  \BibitemOpen
  \bibfield  {author} {\bibinfo {author} {\bibfnamefont {R.}~\bibnamefont
  {Car}}\ and\ \bibinfo {author} {\bibfnamefont {M.}~\bibnamefont {Parrinello}},\
  }\href@noop {} {\bibfield  {journal} {\bibinfo  {journal}
  {\href{http://dx.doi.org/10.1103/PhysRevLett.55.2471}{Phys. Rev. Lett.}}\
  }\textbf {\bibinfo {volume} {55}},\ \bibinfo {pages} {2471} (\bibinfo {year}
  {1985})}\BibitemShut {NoStop}%
\bibitem [{\citenamefont {Perdew}\ \emph
  {et~al.}(1996{\natexlab{b}})\citenamefont {Perdew}, \citenamefont {Burke},\
  and\ \citenamefont {Ernzerhof}}]{perdew_prl_1996}%
  \BibitemOpen
  \bibfield  {author} {\bibinfo {author} {\bibfnamefont {J.~P.}\ \bibnamefont
  {Perdew}}, \bibinfo {author} {\bibfnamefont {K.}~\bibnamefont {Burke}}, \
  and\ \bibinfo {author} {\bibfnamefont {M.}~\bibnamefont {Ernzerhof}},\
  }\href@noop {} {\bibfield  {journal} {\bibinfo  {journal}
  {\href{http://dx.doi.org/10.1103/PhysRevLett.77.3865}{Phys. Rev. Lett.}}\
  }\textbf {\bibinfo {volume} {77}},\ \bibinfo {pages} {3865} (\bibinfo {year}
  {1996}{\natexlab{b}})}\BibitemShut {NoStop}%
\bibitem [{dis()}]{distasio_unpublished}%
  \BibitemOpen
  \href@noop {} {}\bibinfo {note} {{R. A. DiStasio Jr., B. Santra, H.-Y. Ko,
  and R. Car, \textit{to be published}.}}\BibitemShut {Stop}%
\bibitem [{\citenamefont {Landau}\ and\ \citenamefont
  {Lifshitz}(1969)}]{landau_sm_book_1969}%
  \BibitemOpen
  \bibfield  {author} {\bibinfo {author} {\bibfnamefont {L.~D.}\ \bibnamefont
  {Landau}}\ and\ \bibinfo {author} {\bibfnamefont {E.~M.}\ \bibnamefont
  {Lifshitz}},\ }\href@noop {} {\emph {\bibinfo {title} {Statistical
  Physics}}},\ \bibinfo {series} {Course of Theoretical Physics}, Vol.~\bibinfo
  {volume} {5}\ (\bibinfo  {publisher} {Pergamon Press},\ \bibinfo {address}
  {Oxford},\ \bibinfo {year} {1969})\BibitemShut {NoStop}%
\bibitem [{\citenamefont {Morrone}\ and\ \citenamefont
  {Car}(2008)}]{morrone_prl_2008}%
  \BibitemOpen
  \bibfield  {author} {\bibinfo {author} {\bibfnamefont {J.~A.}\ \bibnamefont
  {Morrone}}\ and\ \bibinfo {author} {\bibfnamefont {R.}~\bibnamefont {Car}},\
  }\href@noop {} {\bibfield  {journal} {\bibinfo  {journal}
  {\href{http://dx.doi.org/10.1103/PhysRevLett.101.017801}{Phys. Rev. Lett.}}\
  }\textbf {\bibinfo {volume} {101}},\ \bibinfo {pages} {017801} (\bibinfo
  {year} {2008})}\BibitemShut {NoStop}%
\bibitem [{\citenamefont {Paesani}\ \emph {et~al.}(2007)\citenamefont
  {Paesani}, \citenamefont {Iuchi},\ and\ \citenamefont
  {Voth}}]{paesani_jcp_2007}%
  \BibitemOpen
  \bibfield  {author} {\bibinfo {author} {\bibfnamefont {F.}~\bibnamefont
  {Paesani}}, \bibinfo {author} {\bibfnamefont {S.}~\bibnamefont {Iuchi}}, \
  and\ \bibinfo {author} {\bibfnamefont {G.~A.}\ \bibnamefont {Voth}},\
  }\href@noop {} {\bibfield  {journal} {\bibinfo  {journal}
  {\href{http://dx.doi.org/10.1063/1.2759484}{J. Chem. Phys.}}\ }\textbf
  {\bibinfo {volume} {127}},\ \bibinfo {pages} {074506} (\bibinfo {year}
  {2007})}\BibitemShut {NoStop}%
\bibitem [{\citenamefont {Fanourgakis}\ \emph {et~al.}(2006)\citenamefont
  {Fanourgakis}, \citenamefont {Schenter},\ and\ \citenamefont
  {Xantheas}}]{fanourgakis_jcp_2006}%
  \BibitemOpen
  \bibfield  {author} {\bibinfo {author} {\bibfnamefont {G.~S.}\ \bibnamefont
  {Fanourgakis}}, \bibinfo {author} {\bibfnamefont {G.~K.}\ \bibnamefont
  {Schenter}}, \ and\ \bibinfo {author} {\bibfnamefont {S.~S.}\ \bibnamefont
  {Xantheas}},\ }\href@noop {} {\bibfield  {journal} {\bibinfo  {journal}
  {\href{http://dx.doi.org/10.1063/1.2358137}{J. Chem. Phys.}}\ }\textbf
  {\bibinfo {volume} {125}},\ \bibinfo {pages} {141102} (\bibinfo {year}
  {2006})}\BibitemShut {NoStop}%
\bibitem [{\citenamefont {Giannozzi}\ \emph {et~al.}(2009)\citenamefont
  {Giannozzi}, \citenamefont {Baroni}, \citenamefont {Bonini}, \citenamefont
  {Calandra}, \citenamefont {Car}, \citenamefont {Cavazzoni}, \citenamefont
  {Ceresoli}, \citenamefont {Chiarotti}, \citenamefont {Cococcioni},
  \citenamefont {Dabo}, \citenamefont {{A. Dal Corso}}, \citenamefont {{S. de
  Gironcoli}}, \citenamefont {Fabris}, \citenamefont {Fratesi}, \citenamefont
  {Gebauer}, \citenamefont {Gerstmann}, \citenamefont {Gougoussis},
  \citenamefont {Kokalj}, \citenamefont {Lazzeri}, \citenamefont {{L.
  Martin-Samos}}, \citenamefont {Marzari}, \citenamefont {Mauri}, \citenamefont
  {Mazzarello}, \citenamefont {Paolini}, \citenamefont {Pasquarello},
  \citenamefont {Paulatto}, \citenamefont {Sbraccia}, \citenamefont {Scandolo},
  \citenamefont {Sclauzero}, \citenamefont {Seitsonen}, \citenamefont
  {Smogunov}, \citenamefont {Umari},\ and\ \citenamefont
  {Wentzcovitch}}]{QE-2009}%
  \BibitemOpen
  \bibfield  {author} {\bibinfo {author} {\bibfnamefont {P.}~\bibnamefont
  {Giannozzi}}, \bibinfo {author} {\bibfnamefont {S.}~\bibnamefont {Baroni}},
  \bibinfo {author} {\bibfnamefont {N.}~\bibnamefont {Bonini}}, \bibinfo
  {author} {\bibfnamefont {M.}~\bibnamefont {Calandra}}, \bibinfo {author}
  {\bibfnamefont {R.}~\bibnamefont {Car}}, \bibinfo {author} {\bibfnamefont
  {C.}~\bibnamefont {Cavazzoni}}, \bibinfo {author} {\bibfnamefont
  {D.}~\bibnamefont {Ceresoli}}, \bibinfo {author} {\bibfnamefont {G.~L.}\
  \bibnamefont {Chiarotti}}, \bibinfo {author} {\bibfnamefont {M.}~\bibnamefont
  {Cococcioni}}, \bibinfo {author} {\bibfnamefont {I.}~\bibnamefont {Dabo}},
  \bibinfo {author} {\bibnamefont {{A. Dal Corso}}}, \bibinfo {author}
  {\bibnamefont {{S. de Gironcoli}}}, \bibinfo {author} {\bibfnamefont
  {S.}~\bibnamefont {Fabris}}, \bibinfo {author} {\bibfnamefont
  {G.}~\bibnamefont {Fratesi}}, \bibinfo {author} {\bibfnamefont
  {R.}~\bibnamefont {Gebauer}}, \bibinfo {author} {\bibfnamefont
  {U.}~\bibnamefont {Gerstmann}}, \bibinfo {author} {\bibfnamefont
  {C.}~\bibnamefont {Gougoussis}}, \bibinfo {author} {\bibfnamefont
  {A.}~\bibnamefont {Kokalj}}, \bibinfo {author} {\bibfnamefont
  {M.}~\bibnamefont {Lazzeri}}, \bibinfo {author} {\bibnamefont {{L.
  Martin-Samos}}}, \bibinfo {author} {\bibfnamefont {N.}~\bibnamefont
  {Marzari}}, \bibinfo {author} {\bibfnamefont {F.}~\bibnamefont {Mauri}},
  \bibinfo {author} {\bibfnamefont {R.}~\bibnamefont {Mazzarello}}, \bibinfo
  {author} {\bibfnamefont {S.}~\bibnamefont {Paolini}}, \bibinfo {author}
  {\bibfnamefont {A.}~\bibnamefont {Pasquarello}}, \bibinfo {author}
  {\bibfnamefont {L.}~\bibnamefont {Paulatto}}, \bibinfo {author}
  {\bibfnamefont {C.}~\bibnamefont {Sbraccia}}, \bibinfo {author}
  {\bibfnamefont {S.}~\bibnamefont {Scandolo}}, \bibinfo {author}
  {\bibfnamefont {G.}~\bibnamefont {Sclauzero}}, \bibinfo {author}
  {\bibfnamefont {A.~P.}\ \bibnamefont {Seitsonen}}, \bibinfo {author}
  {\bibfnamefont {A.}~\bibnamefont {Smogunov}}, \bibinfo {author}
  {\bibfnamefont {P.}~\bibnamefont {Umari}}, \ and\ \bibinfo {author}
  {\bibfnamefont {R.~M.}\ \bibnamefont {Wentzcovitch}},\ }\href@noop {}
  {\bibfield  {journal} {\bibinfo  {journal}
  {\href{http://dx.doi.org/10.1088/0953-8984/21/39/395502}{J. Phys.: Condens.
  Matter}}\ }\textbf {\bibinfo {volume} {21}},\ \bibinfo {pages} {395502}
  (\bibinfo {year} {2009})}\BibitemShut {NoStop}%
\bibitem [{\citenamefont {Marzari}\ and\ \citenamefont
  {Vanderbilt}(1997)}]{marzari_prb_1997}%
  \BibitemOpen
  \bibfield  {author} {\bibinfo {author} {\bibfnamefont {N.}~\bibnamefont
  {Marzari}}\ and\ \bibinfo {author} {\bibfnamefont {D.}~\bibnamefont
  {Vanderbilt}},\ }\href@noop {} {\bibfield  {journal} {\bibinfo  {journal}
  {\href{http://dx.doi.org/10.1103/PhysRevB.56.12847}{Phys. Rev. B}}\ }\textbf
  {\bibinfo {volume} {56}},\ \bibinfo {pages} {12847} (\bibinfo {year}
  {1997})}\BibitemShut {NoStop}%
\bibitem [{\citenamefont {Wu}\ \emph {et~al.}(2009)\citenamefont {Wu},
  \citenamefont {Selloni},\ and\ \citenamefont {Car}}]{wu_prb_2009}%
  \BibitemOpen
  \bibfield  {author} {\bibinfo {author} {\bibfnamefont {X.}~\bibnamefont
  {Wu}}, \bibinfo {author} {\bibfnamefont {A.}~\bibnamefont {Selloni}}, \ and\
  \bibinfo {author} {\bibfnamefont {R.}~\bibnamefont {Car}},\ }\href@noop {}
  {\bibfield  {journal} {\bibinfo  {journal}
  {\href{http://dx.doi.org/10.1103/PhysRevB.79.085102}{Phys. Rev. B}}\ }\textbf
  {\bibinfo {volume} {79}},\ \bibinfo {pages} {085102} (\bibinfo {year}
  {2009})}\BibitemShut {NoStop}%
\bibitem [{ko_()}]{ko_unpublished}%
  \BibitemOpen
  \href@noop {} {}\bibinfo {note} {{H.-Y. Ko, B. Santra, R. A. DiStasio Jr., L.
  Kong, Z. Li, X. Wu, and R. Car, ``Enabling Hybrid Density Functional Theory
  Calculations on Large-Scale Condensed-Phase Materials,'' \textit{to be
  published}.}}\BibitemShut {Stop}%
\bibitem [{\citenamefont {Abascal}\ \emph {et~al.}(2005)\citenamefont
  {Abascal}, \citenamefont {Sanz}, \citenamefont {{Garc\'{\i}a
  Fern\'{a}ndez}},\ and\ \citenamefont {Vega}}]{Abascal2005}%
  \BibitemOpen
  \bibfield  {author} {\bibinfo {author} {\bibfnamefont {J.~L.~F.}\
  \bibnamefont {Abascal}}, \bibinfo {author} {\bibfnamefont {E.}~\bibnamefont
  {Sanz}}, \bibinfo {author} {\bibfnamefont {R.}~\bibnamefont {{Garc\'{\i}a
  Fern\'{a}ndez}}}, \ and\ \bibinfo {author} {\bibfnamefont {C.}~\bibnamefont
  {Vega}},\ }\href@noop {} {\bibfield  {journal} {\bibinfo  {journal}
  {\href{http://dx.doi.org/10.1063/1.1931662}{J. Chem. Phys.}}\ }\textbf
  {\bibinfo {volume} {122}},\ \bibinfo {pages} {234511} (\bibinfo {year}
  {2005})}\BibitemShut {NoStop}%
\bibitem [{\citenamefont {Todorov}\ \emph {et~al.}(2006)\citenamefont
  {Todorov}, \citenamefont {Smith}, \citenamefont {Trachenkob},\ and\
  \citenamefont {Dove}}]{Todorov2006}%
  \BibitemOpen
  \bibfield  {author} {\bibinfo {author} {\bibfnamefont {I.~T.}\ \bibnamefont
  {Todorov}}, \bibinfo {author} {\bibfnamefont {W.}~\bibnamefont {Smith}},
  \bibinfo {author} {\bibfnamefont {K.}~\bibnamefont {Trachenkob}}, \ and\
  \bibinfo {author} {\bibfnamefont {M.~T.}\ \bibnamefont {Dove}},\ }\href@noop
  {} {\bibfield  {journal} {\bibinfo  {journal}
  {\href{http://dx.doi.org/10.1039/B517931A}{J. Mater. Chem.}}\ }\textbf
  {\bibinfo {volume} {16}},\ \bibinfo {pages} {1911} (\bibinfo {year}
  {2006})}\BibitemShut {NoStop}%
\bibitem [{\citenamefont {Martyna}\ \emph {et~al.}(1992)\citenamefont
  {Martyna}, \citenamefont {Klein},\ and\ \citenamefont
  {Tuckerman}}]{martyna_jcp_1992}%
  \BibitemOpen
  \bibfield  {author} {\bibinfo {author} {\bibfnamefont {G.~J.}\ \bibnamefont
  {Martyna}}, \bibinfo {author} {\bibfnamefont {M.~L.}\ \bibnamefont {Klein}},
  \ and\ \bibinfo {author} {\bibfnamefont {M.}~\bibnamefont {Tuckerman}},\
  }\href@noop {} {\bibfield  {journal} {\bibinfo  {journal}
  {\href{http://dx.doi.org/10.1063/1.463940}{J. Chem. Phys.}}\ }\textbf
  {\bibinfo {volume} {97}},\ \bibinfo {pages} {2635} (\bibinfo {year}
  {1992})}\BibitemShut {NoStop}%
\bibitem [{\citenamefont {Essmann}\ \emph {et~al.}(1995)\citenamefont
  {Essmann}, \citenamefont {Perera}, \citenamefont {Berkowitz}, \citenamefont
  {Darden}, \citenamefont {Lee},\ and\ \citenamefont {Pedersen}}]{Essmann1995}%
  \BibitemOpen
  \bibfield  {author} {\bibinfo {author} {\bibfnamefont {U.}~\bibnamefont
  {Essmann}}, \bibinfo {author} {\bibfnamefont {L.}~\bibnamefont {Perera}},
  \bibinfo {author} {\bibfnamefont {M.~L.}\ \bibnamefont {Berkowitz}}, \bibinfo
  {author} {\bibfnamefont {T.}~\bibnamefont {Darden}}, \bibinfo {author}
  {\bibfnamefont {H.}~\bibnamefont {Lee}}, \ and\ \bibinfo {author}
  {\bibfnamefont {L.~G.}\ \bibnamefont {Pedersen}},\ }\href@noop {} {\bibfield
  {journal} {\bibinfo  {journal} {\href{http://dx.doi.org/10.1063/1.470117}{J.
  Chem. Phys.}}\ }\textbf {\bibinfo {volume} {103}},\ \bibinfo {pages} {8577}
  (\bibinfo {year} {1995})}\BibitemShut {NoStop}%
\bibitem [{rdf()}]{rdf_footnote}%
  \BibitemOpen
  \href@noop {} {}\bibinfo {note} {{Note that even though the P(I)
  distributions of liquid water obtained from the PBE0+TS-vdW(SC) XC functional
  and the TIP4P/Ice classical force field at 330 K agree well, the
  corresponding $g_{\rm OO}(r)$ show some differences. The largest difference
  found was in the intensity of the first peak of the $g_{\rm OO}(r)$, which
  has a value of 2.51 (PBE0+TS-vdW(SC)) vs. 3.17 (TIP4P/Ice). However, the
  differences beyond the first coordination shell are significantly reduced,
  leading to second and third coordination shells that are in relatively good
  agreement. In this regard, the standard deviation between the intensities of
  these two $g_{\rm OO}(r)$ is only 0.03 when averaged over an oxygen-oxygen
  distance range of 3.5--7.8 \AA.}}\BibitemShut {Stop}%
\bibitem [{\citenamefont {Holten}\ and\ \citenamefont
  {Anisimov}(2012)}]{Holten2012}%
  \BibitemOpen
  \bibfield  {author} {\bibinfo {author} {\bibfnamefont {V.}~\bibnamefont
  {Holten}}\ and\ \bibinfo {author} {\bibfnamefont {M.~A.}\ \bibnamefont
  {Anisimov}},\ }\href@noop {} {\bibfield  {journal} {\bibinfo  {journal}
  {\href{http://dx.doi.org/10.1038/srep00713}{Sci. Rep.}}\ }\textbf {\bibinfo
  {volume} {2}},\ \bibinfo {pages} {713} (\bibinfo {year} {2012})}\BibitemShut
  {NoStop}%
\bibitem [{\citenamefont {Stillinger}\ and\ \citenamefont
  {Weber}(1984{\natexlab{a}})}]{StillingerJCP1984}%
  \BibitemOpen
  \bibfield  {author} {\bibinfo {author} {\bibfnamefont {F.~H.}\ \bibnamefont
  {Stillinger}}\ and\ \bibinfo {author} {\bibfnamefont {T.~A.}\ \bibnamefont
  {Weber}},\ }\href@noop {} {\bibfield  {journal} {\bibinfo  {journal}
  {\href{http://dx.doi.org/10.1063/1.447223}{J. Chem. Phys.}}\ }\textbf
  {\bibinfo {volume} {80}},\ \bibinfo {pages} {4434} (\bibinfo {year}
  {1984}{\natexlab{a}})}\BibitemShut {NoStop}%
\bibitem [{\citenamefont {Stillinger}\ and\ \citenamefont
  {Weber}(1984{\natexlab{b}})}]{Stillinger1984}%
  \BibitemOpen
  \bibfield  {author} {\bibinfo {author} {\bibfnamefont {F.~H.}\ \bibnamefont
  {Stillinger}}\ and\ \bibinfo {author} {\bibfnamefont {T.~A.}\ \bibnamefont
  {Weber}},\ }\href@noop {} {\bibfield  {journal} {\bibinfo  {journal}
  {\href{http://dx.doi.org/10.1126/science.225.4666.983}{Science}}\ }\textbf
  {\bibinfo {volume} {225}},\ \bibinfo {pages} {983} (\bibinfo {year}
  {1984}{\natexlab{b}})}\BibitemShut {NoStop}%
\bibitem [{\citenamefont {Stillinger}(1988)}]{Stillinger1988}%
  \BibitemOpen
  \bibfield  {author} {\bibinfo {author} {\bibfnamefont {F.~H.}\ \bibnamefont
  {Stillinger}},\ }\href@noop {} {\bibfield  {journal} {\bibinfo  {journal}
  {\href{http://dx.doi.org/10.1063/1.454853}{J. Chem. Phys.}}\ }\textbf
  {\bibinfo {volume} {89}},\ \bibinfo {pages} {4180} (\bibinfo {year}
  {1988})}\BibitemShut {NoStop}%
\bibitem [{\citenamefont {Tassone}\ \emph {et~al.}(1994)\citenamefont
  {Tassone}, \citenamefont {Mauri},\ and\ \citenamefont {Car}}]{Tassone1994}%
  \BibitemOpen
  \bibfield  {author} {\bibinfo {author} {\bibfnamefont {F.}~\bibnamefont
  {Tassone}}, \bibinfo {author} {\bibfnamefont {F.}~\bibnamefont {Mauri}}, \
  and\ \bibinfo {author} {\bibfnamefont {R.}~\bibnamefont {Car}},\ }\href@noop
  {} {\bibfield  {journal} {\bibinfo  {journal}
  {\href{http://dx.doi.org/10.1103/PhysRevB.50.10561}{Phys. Rev. B}}\ }\textbf
  {\bibinfo {volume} {50}},\ \bibinfo {pages} {10561} (\bibinfo {year}
  {1994})}\BibitemShut {NoStop}%
\bibitem [{imi()}]{imin_footnote}%
  \BibitemOpen
  \href@noop {} {}\bibinfo {note} {{Here we took the lower limit of the range
  0.14--0.16 \AA$^2$ for $I_{\rm min}$. Any change in the isobestic point
  within that range would produce very minor quantitative
  changes.}}\BibitemShut {Stop}%
\bibitem [{\citenamefont {Marto\v{n}\'{a}k}\ \emph {et~al.}(2004)\citenamefont
  {Marto\v{n}\'{a}k}, \citenamefont {Donadio},\ and\ \citenamefont
  {Parrinello}}]{Martonak2004}%
  \BibitemOpen
  \bibfield  {author} {\bibinfo {author} {\bibfnamefont {R.}~\bibnamefont
  {Marto\v{n}\'{a}k}}, \bibinfo {author} {\bibfnamefont {D.}~\bibnamefont
  {Donadio}}, \ and\ \bibinfo {author} {\bibfnamefont {M.}~\bibnamefont
  {Parrinello}},\ }\href@noop {} {\bibfield  {journal} {\bibinfo  {journal}
  {\href{http://dx.doi.org/10.1103/PhysRevLett.92.225702}{Phys. Rev. Lett.}}\
  }\textbf {\bibinfo {volume} {92}},\ \bibinfo {pages} {225702} (\bibinfo
  {year} {2004})}\BibitemShut {NoStop}%
\bibitem [{\citenamefont {Marto\v{n}\'{a}k}\ \emph {et~al.}(2005)\citenamefont
  {Marto\v{n}\'{a}k}, \citenamefont {Donadio},\ and\ \citenamefont
  {Parrinello}}]{Martonak2005}%
  \BibitemOpen
  \bibfield  {author} {\bibinfo {author} {\bibfnamefont {R.}~\bibnamefont
  {Marto\v{n}\'{a}k}}, \bibinfo {author} {\bibfnamefont {D.}~\bibnamefont
  {Donadio}}, \ and\ \bibinfo {author} {\bibfnamefont {M.}~\bibnamefont
  {Parrinello}},\ }\href@noop {} {\bibfield  {journal} {\bibinfo  {journal}
  {\href{http://dx.doi.org/10.1063/1.1870852}{J. Chem. Phys.}}\ }\textbf
  {\bibinfo {volume} {122}},\ \bibinfo {pages} {134501} (\bibinfo {year}
  {2005})}\BibitemShut {NoStop}%
\bibitem [{hbo()}]{hbond_footnote}%
  \BibitemOpen
  \href@noop {} {}\bibinfo {note} {{In this work, we have chosen to utilize a
  hydrogen bond definition based on the radial distribution functions obtained
  from PBE0+TS-vdW(SC) (330K) liquid water, wherein the defining parameters for
  an intact hydrogen bond between a pair of water molecules include: an
  oxygen-oxygen distance $d_{\rm OO}<3.33$ \AA\ based on the first minimum in
  the $g_{\rm OO}(r)$, an oxygen-hydrogen distance $d_{\rm OH}<2.45$ \AA\ based
  on the second minimum in the $g_{\rm OH}(r)$, and a commonly used hydrogen
  bond angle~\cite{morrone_prl_2008} $\alpha\equiv\angle {\rm O_A \cdots
  H_D\!\!-\!\!O_D}$ $>140^\circ$, where $A$ and $D$ denote the acceptor and
  donor water molecules, respectively. This set of criteria is stricter than
  the widely used hydrogen bond definition of Luzar and
  Chandler.~\cite{luzar_prl_1996} For example, the average number of intact
  hydrogen bonds per water molecule in PBE0+TS-vdW(SC) (330K) liquid
  water~\cite{Distasio2014} was found to be 3.26 with this definition, which is
  smaller than the value of 3.48 computed from the definition of Luzar and
  Chandler. However, the qualitative features reported herein are robust and
  independent of the hydrogen bond definition employed.}}\BibitemShut {Stop}%
\bibitem [{\citenamefont {King}(1967)}]{King1967}%
  \BibitemOpen
  \bibfield  {author} {\bibinfo {author} {\bibfnamefont {S.~V.}\ \bibnamefont
  {King}},\ }\href@noop {} {\bibfield  {journal} {\bibinfo  {journal}
  {\href{http://dx.doi.org/10.1038/2131112a0}{Nature}}\ }\textbf {\bibinfo
  {volume} {213}},\ \bibinfo {pages} {1112} (\bibinfo {year}
  {1967})}\BibitemShut {NoStop}%
\bibitem [{\citenamefont {Fuchs}\ \emph {et~al.}(2008)\citenamefont {Fuchs},
  \citenamefont {Zharnikov}, \citenamefont {Weinhardt}, \citenamefont {Blum},
  \citenamefont {Weigand}, \citenamefont {Zubavichus}, \citenamefont {B\"ar},
  \citenamefont {Maier}, \citenamefont {Denlinger}, \citenamefont {Heske},
  \citenamefont {Grunze},\ and\ \citenamefont {Umbach}}]{Fuchs2008}%
  \BibitemOpen
  \bibfield  {author} {\bibinfo {author} {\bibfnamefont {O.}~\bibnamefont
  {Fuchs}}, \bibinfo {author} {\bibfnamefont {M.}~\bibnamefont {Zharnikov}},
  \bibinfo {author} {\bibfnamefont {L.}~\bibnamefont {Weinhardt}}, \bibinfo
  {author} {\bibfnamefont {M.}~\bibnamefont {Blum}}, \bibinfo {author}
  {\bibfnamefont {M.}~\bibnamefont {Weigand}}, \bibinfo {author} {\bibfnamefont
  {Y.}~\bibnamefont {Zubavichus}}, \bibinfo {author} {\bibfnamefont
  {M.}~\bibnamefont {B\"ar}}, \bibinfo {author} {\bibfnamefont
  {F.}~\bibnamefont {Maier}}, \bibinfo {author} {\bibfnamefont {J.~D.}\
  \bibnamefont {Denlinger}}, \bibinfo {author} {\bibfnamefont {C.}~\bibnamefont
  {Heske}}, \bibinfo {author} {\bibfnamefont {M.}~\bibnamefont {Grunze}}, \
  and\ \bibinfo {author} {\bibfnamefont {E.}~\bibnamefont {Umbach}},\
  }\href@noop {} {\bibfield  {journal} {\bibinfo  {journal}
  {\href{http://dx.doi.org/10.1103/PhysRevLett.100.027801}{Phys. Rev. Lett.}}\
  }\textbf {\bibinfo {volume} {100}},\ \bibinfo {pages} {027801} (\bibinfo
  {year} {2008})}\BibitemShut {NoStop}%
\bibitem [{\citenamefont {Tokushima}\ \emph {et~al.}(2008)\citenamefont
  {Tokushima}, \citenamefont {Harada}, \citenamefont {Takahashi}, \citenamefont
  {Senba}, \citenamefont {Ohashi}, \citenamefont {Pettersson}, \citenamefont
  {Nilsson},\ and\ \citenamefont {Shin}}]{Tokushima2008}%
  \BibitemOpen
  \bibfield  {author} {\bibinfo {author} {\bibfnamefont {T.}~\bibnamefont
  {Tokushima}}, \bibinfo {author} {\bibfnamefont {Y.}~\bibnamefont {Harada}},
  \bibinfo {author} {\bibfnamefont {O.}~\bibnamefont {Takahashi}}, \bibinfo
  {author} {\bibfnamefont {Y.}~\bibnamefont {Senba}}, \bibinfo {author}
  {\bibfnamefont {H.}~\bibnamefont {Ohashi}}, \bibinfo {author} {\bibfnamefont
  {L.}~\bibnamefont {Pettersson}}, \bibinfo {author} {\bibfnamefont
  {A.}~\bibnamefont {Nilsson}}, \ and\ \bibinfo {author} {\bibfnamefont
  {S.}~\bibnamefont {Shin}},\ }\href@noop {} {\bibfield  {journal} {\bibinfo
  {journal} {\href{http://dx.doi.org/10.1016/j.cplett.2008.04.077}{Chemical
  Physics Letters}}\ }\textbf {\bibinfo {volume} {460}},\ \bibinfo {pages} {387
  } (\bibinfo {year} {2008})}\BibitemShut {NoStop}%
\bibitem [{\citenamefont {Nilsson}\ \emph {et~al.}(2010)\citenamefont
  {Nilsson}, \citenamefont {Nordlund}, \citenamefont {Waluyo}, \citenamefont
  {Huang}, \citenamefont {Ogasawara}, \citenamefont {Kaya}, \citenamefont
  {Bergmann}, \citenamefont {N\"{a}slund}, \citenamefont {\"{O}str\"{o}m},
  \citenamefont {Wernet}, \citenamefont {Andersson}, \citenamefont {Schiros},\
  and\ \citenamefont {Pettersson}}]{Nilsson2010}%
  \BibitemOpen
  \bibfield  {author} {\bibinfo {author} {\bibfnamefont {A.}~\bibnamefont
  {Nilsson}}, \bibinfo {author} {\bibfnamefont {D.}~\bibnamefont {Nordlund}},
  \bibinfo {author} {\bibfnamefont {I.}~\bibnamefont {Waluyo}}, \bibinfo
  {author} {\bibfnamefont {N.}~\bibnamefont {Huang}}, \bibinfo {author}
  {\bibfnamefont {H.}~\bibnamefont {Ogasawara}}, \bibinfo {author}
  {\bibfnamefont {S.}~\bibnamefont {Kaya}}, \bibinfo {author} {\bibfnamefont
  {U.}~\bibnamefont {Bergmann}}, \bibinfo {author} {\bibfnamefont {L.~A.}\
  \bibnamefont {N\"{a}slund}}, \bibinfo {author} {\bibfnamefont
  {H.}~\bibnamefont {\"{O}str\"{o}m}}, \bibinfo {author} {\bibfnamefont
  {P.}~\bibnamefont {Wernet}}, \bibinfo {author} {\bibfnamefont {K.~J.}\
  \bibnamefont {Andersson}}, \bibinfo {author} {\bibfnamefont {T.}~\bibnamefont
  {Schiros}}, \ and\ \bibinfo {author} {\bibfnamefont {L.~G.~M.}\ \bibnamefont
  {Pettersson}},\ }\href@noop {} {\bibfield  {journal} {\bibinfo  {journal}
  {\href{http://dx.doi.org/10.1016/j.elspec.2010.02.005}{J. Electron Spectros.
  Relat. Phenomena}}\ }\textbf {\bibinfo {volume} {177}},\ \bibinfo {pages}
  {99} (\bibinfo {year} {2010})}\BibitemShut {NoStop}%
\bibitem [{\citenamefont {Skone}\ \emph {et~al.}(2014)\citenamefont {Skone},
  \citenamefont {Govoni},\ and\ \citenamefont {Galli}}]{Skone2014}%
  \BibitemOpen
  \bibfield  {author} {\bibinfo {author} {\bibfnamefont {J.~H.}\ \bibnamefont
  {Skone}}, \bibinfo {author} {\bibfnamefont {M.}~\bibnamefont {Govoni}}, \
  and\ \bibinfo {author} {\bibfnamefont {G.}~\bibnamefont {Galli}},\
  }\href@noop {} {\bibfield  {journal} {\bibinfo  {journal}
  {\href{http://dx.doi.org/10.1103/PhysRevB.89.195112}{Phys. Rev. B}}\ }\textbf
  {\bibinfo {volume} {89}},\ \bibinfo {pages} {195112} (\bibinfo {year}
  {2014})}\BibitemShut {NoStop}%
\bibitem [{\citenamefont {Tkatchenko}\ \emph {et~al.}(2012)\citenamefont
  {Tkatchenko}, \citenamefont {{R. A. DiStasio Jr.}}, \citenamefont {Car},\
  and\ \citenamefont {Scheffler}}]{tkatchenko_prl_2012}%
  \BibitemOpen
  \bibfield  {author} {\bibinfo {author} {\bibfnamefont {A.}~\bibnamefont
  {Tkatchenko}}, \bibinfo {author} {\bibnamefont {{R. A. DiStasio Jr.}}},
  \bibinfo {author} {\bibfnamefont {R.}~\bibnamefont {Car}}, \ and\ \bibinfo
  {author} {\bibfnamefont {M.}~\bibnamefont {Scheffler}},\ }\href@noop {}
  {\bibfield  {journal} {\bibinfo  {journal}
  {\href{http://dx.doi.org/10.1103/PhysRevLett.108.236402}{Phys. Rev. Lett.}}\
  }\textbf {\bibinfo {volume} {108}},\ \bibinfo {pages} {236402} (\bibinfo
  {year} {2012})}\BibitemShut {NoStop}%
\bibitem [{\citenamefont {{R. A. DiStasio Jr.}}\ \emph
  {et~al.}(2012)\citenamefont {{R. A. DiStasio Jr.}}, \citenamefont {{O. A. von
  Lilienfeld}},\ and\ \citenamefont {Tkatchenko}}]{distasio_pnas_2012}%
  \BibitemOpen
  \bibfield  {author} {\bibinfo {author} {\bibnamefont {{R. A. DiStasio Jr.}}},
  \bibinfo {author} {\bibnamefont {{O. A. von Lilienfeld}}}, \ and\ \bibinfo
  {author} {\bibfnamefont {A.}~\bibnamefont {Tkatchenko}},\ }\href@noop {}
  {\bibfield  {journal} {\bibinfo  {journal}
  {\href{http://dx.doi.org/10.1073/pnas.1208121109}{Proc. Natl. Acad. Sci.
  USA}}\ }\textbf {\bibinfo {volume} {109}},\ \bibinfo {pages} {14791}
  (\bibinfo {year} {2012})}\BibitemShut {NoStop}%
\bibitem [{\citenamefont {Ambrosetti}\ \emph {et~al.}(2014)\citenamefont
  {Ambrosetti}, \citenamefont {Reilly}, \citenamefont {{R. A. DiStasio Jr.}},\
  and\ \citenamefont {Tkatchenko}}]{ambrosetti_jcp_2014}%
  \BibitemOpen
  \bibfield  {author} {\bibinfo {author} {\bibfnamefont {A.}~\bibnamefont
  {Ambrosetti}}, \bibinfo {author} {\bibfnamefont {A.~M.}\ \bibnamefont
  {Reilly}}, \bibinfo {author} {\bibnamefont {{R. A. DiStasio Jr.}}}, \ and\
  \bibinfo {author} {\bibfnamefont {A.}~\bibnamefont {Tkatchenko}},\
  }\href@noop {} {\bibfield  {journal} {\bibinfo  {journal}
  {\href{http://dx.doi.org/10.1063/1.4865104}{J. Chem. Phys.}}\ }\textbf
  {\bibinfo {volume} {140}},\ \bibinfo {pages} {18A508} (\bibinfo {year}
  {2014})}\BibitemShut {NoStop}%
\bibitem [{\citenamefont {{R. A. DiStasio Jr.}}\ \emph
  {et~al.}(2014)\citenamefont {{R. A. DiStasio Jr.}}, \citenamefont {Gobre},\
  and\ \citenamefont {Tkatchenko}}]{distasio_jpcm_2014}%
  \BibitemOpen
  \bibfield  {author} {\bibinfo {author} {\bibnamefont {{R. A. DiStasio Jr.}}},
  \bibinfo {author} {\bibfnamefont {V.~V.}\ \bibnamefont {Gobre}}, \ and\
  \bibinfo {author} {\bibfnamefont {A.}~\bibnamefont {Tkatchenko}},\
  }\href@noop {} {\bibfield  {journal} {\bibinfo  {journal}
  {\href{http://dx.doi.org/10.1088/0953-8984/26/21/213202}{J. Phys.: Condens.
  Matter}}\ }\textbf {\bibinfo {volume} {26}},\ \bibinfo {pages} {213202}
  (\bibinfo {year} {2014})}\BibitemShut {NoStop}%
\bibitem [{\citenamefont {Luzar}\ and\ \citenamefont
  {Chandler}(1996)}]{luzar_prl_1996}%
  \BibitemOpen
  \bibfield  {author} {\bibinfo {author} {\bibfnamefont {A.}~\bibnamefont
  {Luzar}}\ and\ \bibinfo {author} {\bibfnamefont {D.}~\bibnamefont
  {Chandler}},\ }\href@noop {} {\bibfield  {journal} {\bibinfo  {journal}
  {\href{http://dx.doi.org/10.1103/PhysRevLett.76.928}{Phys. Rev. Lett.}}\
  }\textbf {\bibinfo {volume} {76}},\ \bibinfo {pages} {928} (\bibinfo {year}
  {1996})}\BibitemShut {NoStop}%
\end{thebibliography}

%

\end{document}